\documentclass[12pt,preprint]{aastex}
\begin{document}


\title{A Keck/HIRES\altaffilmark{1} Study of Kinematics of the Cold
  Interstellar Medium in Dwarf Starburst Galaxies.}   
 
\author{Colleen M. Schwartz\altaffilmark{2} and
  Crystal L. Martin\altaffilmark{3,4,5} }
\affil{Department of Physics, University of California, Santa Barbara,
CA 93106}

\altaffiltext{1}{Data presented herein were obtained at the
  W.M. Keck Observatory, which is operated as a scientific partnership
  among the California Institute of Technology, the University of
  California and the National Aeronautics and Space
  Administration. The Observatory was made possible by the generous
  financial support of the W.M. Keck Foundation.} 
\altaffiltext{2}{colleen@physics.ucsb.edu}
\altaffiltext{3}{cmartin@physics.ucsb.edu}
\altaffiltext{4}{Packard Fellow}
\altaffiltext{5}{Alfred P. Sloan Foundation Fellow}

\shorttitle{\sc Na D Absorption in Dwarf Starburst Galaxies}
\shortauthors{\sc Schwartz \& Martin}

\received{}
\accepted{}

\begin{abstract}
We have obtained high resolution Echelle spectra (R = 30,000 - 50,000)
of the Na D absorption doublet ($\lambda\lambda$5890, 5896) for  
six dwarf starburst galaxies and two more luminous starbursts:  M82
and NGC 1614.  The absorption features were separated into multiple
components and separated into stellar and interstellar parts based on
kinematics. We find that three of the dwarfs show outflows, with an
average blueshift of 27 km s$^{-1}$.  This is small compared to the
highest velocity components in NGC 1614 and M82 (blueshifted by 150 km
s$^{-1}$ and 91 km s$^{-1}$, respectively); these two brighter
galaxies also show more complex absorption profiles than the
dwarfs. None of the outflow speeds clearly exceed the escape velocity
of the host galaxy.  Sightlines in NGC 2363 and NGC4214 apparently
intersect expanding shells.  We compare the shocked gas velocity
(v$_{NaD}$) to the ionized gas velocity (v$_{H\alpha}$) and interpret
the velocity difference as either a trapped ionization front (NGC
4214) or a leaky \ion{H}{2} region (NGC 2363).  The dwarfs show
N$_{NaD}$ = 10$^{11.8-13.7}$ cm$^{-2}$, while the Na D columns in M82
and NGC 1614 are 10$^{13.7}$ cm$^{-2}$ and 10$^{14.0}$ cm$^{-2}$,
respectively.  The mass of expelled gas is highly sensitive to outflow
geometry, dust depletion, and ionization fraction, but with a simple
shell model we estimate neutral outflow gas masses from $\sim$10$^6$
M$_\odot$ to $\sim$10$^{10}$ M$_\odot$.   

\end{abstract}




\section{Introduction}

Galactic winds are driven by supernovae and stellar winds in
starburst galaxies.  They drive outflows of metals and dust
from the interstellar medium (ISM) into the galactic halo and even
into the intergalactic medium (IGM).  These outflows may also heat
the intracluster medium and enrich it with metals, although the amount
of matter permanently escaping the galaxy is still debated.  The
amount of interstellar gas expelled into the IGM by a starburst in a
``blowout'' is thought to depend on the star formation rate within the
galaxy, the escape velocity of the galaxy, and the presence of an
extended, low-density gaseous halo (Silich \& Tenorio-Tagle 1998;
Legrand et al. 2001).  In dwarf starbursts, the mass-loss measurements
have focused on the ionized components of this multi-phase flow.  In
this paper, we explore the kinematics of the cold, neutral clouds in
dwarf starburst outflows. 

Due to their low gravitational potential, dwarf starburst galaxies
have small binding energies and may be particularly vulnerable to
large mass loss via superwinds.  Previous works have studied X-ray
emission from hot gas within the starbursts (Dahlem, Weaver, \&
Heckman 1998; Martin, Kobulnicky, \& Heckman 2002), and optical line
emission (e.g. Lehnert \& Heckman 1996) and absorption (Martin \&
Armus 2004; Rupke, Vielleux, \& Sanders 2002; Heckman et al. 2000 -
hereafter HLSA) from larger, more luminous galaxies. Other studies
have observed dwarf galaxies in emission (Marlowe et al. 1995; Martin
1999), but ours is among the first to study dwarf starbursts in
neutral absorption lines at optical wavelengths.  We present a small
sample of dwarf starburst galaxies and attempt to quantify mass flux
in the cold neutral medium (CNM).  

The advantage of looking at absorption lines is that we can eliminate
all ambiguity in the direction of the radial flow of gas.  Using the
starburst itself as a background continuum source, it is known
absolutely whether the gas we see in absorption is falling in
(redshifted) or expanding outward (blueshifted). Moreover, unlike
emission lines, whose strength scales as density squared, the strength
of spectral lines seen in absorption are directly proportional to the
column density of the neutral gas along the line of sight.   Several
previous studies (Phillips 1993;  Lequeux et al. 1995;
Gonzalez-Delgado et al. 1998; HLSA; Martin \& Armus 2004) have
detected interstellar absorption lines in starburst galaxies of
various sizes and morphologies, finding blueshifted neutral gas and
related large-scale outflows of metals.  HLSA find an average outflow
speed of $\sim$100 km s$^{-1}$ for luminous infrared galaxies (LIRGs)
and more recently, Martin \& Armus (2004) and  Rupke et al. (2002)
find outflow speeds of up to 700 km s$^{-1}$ for two samples of
ultraluminous infrared galaxies (ULIRGs).   

We focus our observations on the \ion{Na}{1} absorption
doublet (5889.95~\AA\, and  5895.92~\AA, also called the Na D doublet)
because it is stronger than any other optical resonance line
(e.g. \ion{K}{1}, \ion{Ca}{2}); most other cosmically abundant
resonance lines are produced in the UV and therefore cannot be
observed in nearby galaxies with ground-based telescopes.  Na D is a
good tracer of cold neutral gas because its ionization potential (5.14
eV) is less than that of hydrogen, so we can be sure any \ion{Na}{1}
absorption is occuring in a cold region of the ISM.  It is also
important to note that sodium is not as strongly depleted in diffuse,
low-velocity clouds as \ion{K}{1} or \ion{Ca}{2} (Spitzer 1968; Savage
\& Sembach 1996).   By using high resolution spectroscopy we can
resolve individual, kinematically distinct doublet components within
the absorption profiles. At lower resolution, deconvolving the
absorption components of disparate origins is a more difficult task
(see, e.g., HLSA).    

It is our goal to observationally determine the kinematics of the CNM
in our sample of starburst galaxies.  In the second section of this
paper, we discuss the observations, data reduction, and basic data
analysis.  In \S3 we detail our method of separating stellar and
interstellar sodium, discuss the kinematics and widths of the
absorption features, and then use the interstellar sodium to determine
a column density.  In \S4 we present our interpretation of the data,
including simple models for outflow scenarios in the individual
galaxies.  The final section summarizes our major results. 


\section{Observations \& Data Reduction}

We observed eight nearby galaxies using the Keck I High Resolution
Echelle Spectrometer (HIRES; Vogt et al. 1994) on 3-5 February 2000.  
Table 1 lists relevant properties of the galaxies.  Six of the eight
galaxies in the sample are classified as dwarfs; we define dwarf
galaxies to have M$_B \ge$ -18 for our choice of H$_0$ = 75 km s$^{-1}$
Mpc$^{-1}$.  NGC 1614 is a LIRG (Alonso-Herrero et al. 2001).  M82 is
slightly brighter than a dwarf galaxy, but less luminous than a LIRG,
making it a good object with which to study the transition between the
dwarfs and the LIRGs (Sofue 1997). The dwarf galaxies were selected
mainly for compatibility with Martin 1998, which provides reference
H$\alpha$ ($\lambda$6563) narrowband images as well as preselecting
galaxies with intense star formation over the range -13.5 $<$ M$_B$
$<$ -18.5.  The sample of galaxies observed here are also seen in
H$\alpha$ to contain expanding supershells of ionized gas.  NGC 1614
was not observed in that study, but does have large-scale H$\alpha$
structure (Neff et al. 1990). The slit in Martin's 1998 observations
missed the bubble in NGC 5253; however, this galaxy was seen to
contain an expanding shell in H$\alpha$ by Marlowe et al. (1995).     

The galaxies were observed with HIRES using both the red and blue
cross dispersers to obtain full coverage of the desired wavelength
range (i.e. from \ion{Ca}{2} to \ion{K}{1}). The red cross disperser
covered the spectral range 5400~\AA \,--\,7700~\AA;  the blue
cross-disperser produced spectra in the range 3700~\AA~--~5900~\AA,
with Echelle orders overlapping in wavelength to give complete
spectral coverage to $\sim$5000~\AA. One galaxy, NGC 4214,
was observed at two positions, hereafter called NGC 4214-1 and NGC
4214-2 (a bright star cluster at the galactic nucleus and a cluster
of star-forming knots, respectively -- see \S4.2.4).  NGC 4214-2 was
not observed with the blue cross-disperser. In all galaxies, the slit
covered the brightest optical cluster to give the strongest continuum
source against which to measure the absorption.  Decker A12 was used;
the slit size was 10'' $\times$ 0.73'' for NGC 1614 and NGC 4449, and
10'' $\times$ 1.24'' for all other galaxies.

The spectra were processed with the ECHELLE package
in IRAF\footnote{IRAF (Image Reduction and Analysis Facility) is
  distributed by the National Optical Astronomy Observatories, which
  are operated by AURA, Inc., under cooperative agreement with the
  National Science Foundation.} 
using standard techniques.  The CCD chip overscan region was used to
bias-subtract the images, and a quartz lamp was used to flat-field the
images in CCDPROC.  Cosmic rays were removed from the CCD images by
averaging multiple images of the same galaxy taken in succession on
the same night; if multiple frames were unavailable, the IRAF task
COSMICRAY was used to remove singular, bright pixels from the
image. We did not flux-calibrate the spectra, as we are concerned
primarily with measuring the equivalent width of absorption lines,
which are purely a function of the normalized continuum intensity
(Spitzer 1978). A Thorium-Argon lamp was observed for spectral
wavelength calibration, resulting in rms residuals in  each echelle
order of less than 0.5 km s$^{-1}$. The resolution of the spectrum is
11 km s$^{-1}$ (6 km s$^{-1}$) for the 1.24'' (0.73'') slit, as
measured from night sky lines. After the orders were extracted and the
sky background was subtracted using the task DOECSLIT, the spectrum
for each order was divided by the continuum to provide a normalized
spectrum.  The absorption line profiles were fit with Gaussian
components using the IRAF task SPECFIT (Kriss 1994). This allowed us
to fit multiple pairs of sodium lines to each profile and accurately
measure the equivalent width of the lines.  


\section{Results}

The Na D doublet was detected in six of the eight galaxies: NGC 1569,
NGC 1614, NGC 4214-1 and -2, NGC 4449, NGC 5253, and M82.  No absorption
was seen in NGC 2363 or I Zw 18.  We believe the Na D in NGC 5253,
NGC 4214-1, and specific components of NGC 1569 and NGC 4449 are likely 
stellar in nature, via arguments discussed in section 3.1.
The spectra of the five galaxies in which interstellar sodium
absorption was detected are shown in Figure 1; there are a total of 
ten interstellar Na D absorption components.

It is interesting to compare the absorption line kinematics to those
of emission-line gas.  H$\alpha$ kinematics were previously measured
across large regions in the dwarfs in this paper and M82 to map out
superbubbles by Martin 1998. Our observations directly measured the
H$\alpha$ also -- albeit over a much smaller region.  Nonetheless, we
can use the H$\alpha$ along these particular sightlines to tie Na D
kinematics into these larger structures.  The detailed results of
these spectra will be discussed individually in \S4.2


\subsection{Stellar and Interstellar Sodium}

In order to make a statement about the properties of the interstellar
sodium in these galaxies, we need to first determine that the sodium
we see is actually interstellar.  The Na D resonance doublet is
prominent in the spectra of cooler stars.  The spectrum of a dwarf
will be dominated by K-type giants and supergiants if the starburst is
older than $\sim$10 Myr (Leitherer et al. 1999); these stars show
strong photospheric sodium absorption.  Therefore, we must find a way
of deciphering the nature of the neutral sodium observed. 

We have made use of the available stellar spectral lines in our data
to help distinguish between stellar and interstellar sodium. In
particular, the Mg b band triplet (5167.32~\AA, 5172.68~\AA, and
5183.60~\AA) is a good indicator of the presence of certain stellar
populations.  The Mg b triplet is present in F through M stars, with
maximum absorption in the range K0 to M3, and is not present in the
CNM because it is a highly excited line.  Therefore, we can be certain
that any Mg b absorption we see is from a stellar atmosphere.  The
equivalent width of this absorption feature is also well correlated
with that of the stellar Na D absorption; we can use the Mg b to
disjoin the stellar and interstellar sodium by fitting the Na D
absorption profiles with multiple components, one of which corresponds
to a purely stellar feature.  

Using the stellar atlas of Jacoby, Hunter, \& Christian (1984), we
have measured the equivalent widths of the Mg b and Na D absorption
lines in a sample of stars and fit the data with a least-squares fit.
We find
\begin{equation}
\mbox{W(NaD)} = (0.40 \pm 0.05)\times
\mbox{W(Mgb)} + (0.65 \pm 0.28).
\end{equation}
This is shown with the data from the stellar atlas in Figure 2. This
is consistent with the calculations of Rupke et al. (2002) that W(NaD)
$\sim$ 0.5W(Mgb).  We can use this relationship to predict the
strength, width and velocity of the stellar component of the Na D
absorption. There are two galaxies in our sample, NGC 1569 and NGC
4449, which have both a stellar and interstellar component.  NGC 5253
has what appears to be a purely stellar component, and therefore is
not included in the sample of galaxies showing interstellar neutral
sodium absorption. The equivalent widths of the stellar and
interstellar Na D and Mg b for these three galaxies are shown in
Figure 3, and the \ion{Mg}{1} spectra are presented in Figure 4.  The
central component of the M82 absorption profile is near the systemic
velocity (it is redshifted by only 4 km s$^{-1}$); however, no Mg b
absorption is seen to a level of 0.24~\AA.  Since the equivalent width
of the Na D absorption in this central component is 1.44~\AA, we can
safely say that there is no significant population of cool giants and
supergiants contaminating the spectrum at this position, and we will
assume this component is purely interstellar.   

HLSA classify their ``interstellar dominated'' sample galaxies as
having a larger line width than their ``strong stellar contamination''
galaxies. Since our spectra are well resolved and we are able to
distingush between stellar and interstellar lines, it is interesting
to note that we can use this method, rather than the overall width of
the absorption profile, to categorize the Na D.   There are only two
galaxies (NGC 1569 and NGC 4449) where we have both stellar and
interstellar lines, and both of these absorption line systems have a
larger interstellar width than stellar.  Similarly, HLSA found that
systems with largers widths are interstellar dominated. 


\subsection{Kinematics of the Interstellar \ion{Na}{1}}

To discuss the kinematics of the CNM, it is imperative to know the
systemic velocity (v$_{sys}$) of the galaxy, i.e. the heliocentric
velocity of the starburst.  However, one may wish to compare data with
previous observations in which the velocity of the galactic center of
mass was used.  Usually the center-of-mass velocity and the stellar
velocity are similar, unless the galaxy is inclined and the starburst
region is far from the axis of rotation; this could create a velocity
offset between the starburst velocity and that of the galactic
nucleus.  In this case, it may be better to use the velocity of the
stars themselves.  We were able to get stellar velocities for three
galaxies (NGC 1569, NGC 4449, and NGC 5253) by using stellar
absorption lines (see \S3.1). In the other five galaxies, we set the
systemic velocity using (in order of preference) CO maps (NGC 1614,
NGC 4214, and M82), optical lines (NGC 2363), and stellar velocities
(I Zw 18). Table 1 gives the systemic velocities used in this analysis
and their sources. A heliocentric velocity scale is used throughout
the paper. 

The HLSA sample of 32 LIRGs showed significant blueshifts in the Na D
doublet for their ``interstellar dominated'' outflows, indicating
nuclear outflows with typical velocities of $\sim$100 km
s$^{-1}$.  Our data for the dwarf galaxies also show the Na D 
absorption to be blueshifted, although with overall slower outflow
speeds ($\sim$30 km s$^{-1}$).  NGC 1614 and M82 show both redshifted
and blueshifted components; these are further discussed in \S4.2.  The
velocities for individual interstellar absorption components are
given in Table 2.   

The line width (FWHM) of the interstellar components for the dwarfs
and M82 ranges from $\sim$30 to $\sim$80 km s$^{-1}$, with an average
of 43 km s$^{-1}$.  The existence of such broad absorption lines in
neutral sodium leads us to conclude that there are multiple clouds of
cold gas producing each component of the absorption; the thermal width
of a cloud of cold gas at $\sim$100 K would be only $\sim$0.20 km
s$^{-1}$. Moreover, the line widths of cold clouds in the Galactic
halo are found to be just a few km s$^{-1}$ (Spitzer \& Fitzpatrick
1995).  The individual line widths in NGC 1614 are significantly
larger ($\sim$100 km s$^{-1}$ or more) than those in the dwarfs or
M82 ($\sim$45 km s$^{-1}$).  The total width of all components in M82
is relatively large (236 km s$^{-1}$).   

Among the dwarfs in the sample, we do not see a correlation between
the FWHM of the lines and the rest-frame equivalent width.  However,
when the data for M82 and NGC 1614 are added, there is a trend wherein
equivalent width increases in proportion to FWHM, for these N points.  
Given the lack of correlation seen by HLSA using a much
larger data set, we conclude only that the largest values of FWHM (and
similarly the largest equivalent widths) are much smaller in dwarfs
than in brighter galaxies.  


\subsection{Column Densities}

The doublet lines are fit with Gaussian profiles in pairs with SPECFIT
to measure the velocity, FWHM, and equivalent width W of
each line (see Table 2).  If the absorbing gas is optically thin, then
the blue ($\lambda$5890) line will be twice as strong as the red
($\lambda$5896) line, and the ``doublet ratio,'' W$_{blue}$/W$_{red}$
= 2.  If the gas is optically thick, W$_{blue}$/W$_{red}$ approaches
unity as the optical depth $\tau$ approaches infinity.  In our data,
just one component is optically thin, the rest are optically thick.   

We can easily measure how much of the continuum source is being
obscured by absorbing gas clouds by studying the residual normalized
intensity I$_\lambda$ of a line at wavelength $\lambda$. If
I$_\lambda$ = 0 for the blue line of a doublet (i.e. the stronger
line), then no continuum radiation is ``leaking'' into the absorption
line.  In other words, the absorbing cloud completely covers the
continuum source and the covering factor C$_f$ = 1.    If a cloud does
not cover the whole source, the continuum radiation will ``leak'' into
the absorption profile and increase the minimum flux, I$_\lambda$, and
C$_f$ $<$ 1.  Specifically, following the derivation of Barlow \&
Sargent (1997), we can find the covering factor:
\begin{equation}
\mbox{C}_f = \frac{I_{5896}^2 - 2I_{5896} + 1}{I_{5890} - 2I_{5896} + 1}.
\end{equation}
In our data, the covering factor ranges from 0.23 to 1.0 (see Table 2).

Now that we know the covering factor of the absorbing gas, we can
determine the Na D column density.  The curve of growth relates the
optical depth at line center, $\tau_0$, to the doublet ratio via a
function F($\tau_0$), determined numerically (Spitzer 1978).  The
column density is then
\begin{equation}
\mbox{N} = \Big\{ \frac{\pi^{1/2} \tau_0}{2F(\tau_0)} \Big\} 
1.13 \times 10^{20} \frac{\mbox{W}_\lambda}{\lambda^2 f}
\mbox{  cm}^{-2} 
\end{equation}
(Spitzer 1968), where $f$ is the oscillator strength (Morton
1991), and W$_\lambda$ and $\lambda$ are given in Angstroms.

Using this method, we find that N(NaD) in the dwarfs
ranges from 6.2 $\times$ $10^{11}$ cm$^{-2}$ in NGC 1569 to 5.1
$\times$ $10^{13}$ cm$^{-2}$ in NGC 4449.  We find the total neutral
sodium column densities in NGC 1614 and M82 to be 1.0 $\times$
$10^{14}$ cm$^{-2}$ and 2.4 $\times$ $10^{13}$ cm$^{-2}$,
respectively.   The measured equivalent widths, and the calculated
covering factors and Na D column densities, are presented in
Table 2.   From the column density of sodium, we can then use the
metallicity of the galaxy (given in Table 2; (Na/H)$_\odot$ =
2.04$\times 10^{-6}$ -- Martin \& Zalubas 1981) to find the total 
\ion{H}{1} column probed.  A source of systematic uncertainty is the
contingence of neutral sodium column density on the ionization
parameter, $f_{ion}$, as well as the fractional depletion of sodium
onto grains, $f_D$.  Since dwarfs typically have a low metallicity, we
expect lower dust content and depletion level than in larger
galaxies. 
It is the product of these, $f_Df_{ion}$, which is the important
factor in our calculations, as is seen in Wakker \& Mathis (2000).
They measure the abundance of neutral sodium in Galactic high- and
intermediate-velocity clouds.  Using their results and a typical
\ion{H}{1} column of 10$^{20}$ cm$^{-2}$, we calculate $f_Df_{ion}$ =
$\sim$300.  We therefore
parameterize the column densities by a simple factor of
($f_Df_{ion}$/100) to reflect the typical correction, and calculate
the total \ion{H}{1} column using the equation  
\begin{equation}
\frac{\mbox{N(\ion{Na}{1})}}{\mbox{N(\ion{H}{1})}} = \Big( \frac{Na}{H}
\Big)_\odot \Big( \frac{Z}{Z_\odot} \Big) \frac{1}{f_Df_{ion}}.
\end{equation}
We find that the corresponding N(\ion{H}{1}) in the dwarfs could be as
large as 10$^{20}$($f_Df_{ion}/100$) cm$^{-2}$ (NGC 1569) to
10$^{22}$($f_Df_{ion}/100$) cm$^{-2}$ (NGC 4449).  In M82 and NGC
1614, this conversion gives a total \ion{H}{1} column of
10$^{21}$($f_Df_{ion}/100$) cm$^{-2}$ and 10$^{22}$($f_Df_{ion}/100$)
cm$^{-2}$, respectively. 


\subsection{Comparison to \ion{K}{1} Absorption Spectra}

In addition to Na D, the \ion{K}{1} resonance absorption line
($\lambda\lambda 7664.91, 7698.97$) is also present in two of our
spectra.  \ion{K}{1} and \ion{Na}{1} have similar ionization
potentials (4.34 eV and 5.14 eV, respectively) and similar depletion
levels (Savage \& Sembach 1991, Wakker \& Mathis 2000), so it is
expected that the species will be spatially coincident in the CNM.  It
is unfortunate that in nearby galaxies, \ion{K}{1} is found in
spectral regions that are strongly contaminated by the Fraunhofer
A-band absorption from atmospheric O$_2$ (near 7600~\AA). In only two
galaxies were we able to discern the \ion{K}{1}  absorption: NGC 1569
and NGC 1614.  NGC 1569 is luckily blueshifted slightly away from the
sky absorption, while NGC 1614 is  redshifted out of the contaminated
region entirely.  Other galaxies are either completely contaminated by
the Fraunhofer absorption, the \ion{K}{1} doublet is too weak to
measure, or the lines lie between Echelle orders.  

The \ion{K}{1} profile is arguably a better constraint on the cold gas
kinematics than the Na D profile because the doublet spacing is wider,
and the \ion{K}{1} oscillator strength is lower so the lines are less
saturated.  We fit the \ion{K}{1} profile in NGC 1614 with SPECFIT and
found components at -149 km s$^{-1}$ and +70 km s$^{-1}$.  This result
helped us pin down the best component velocities for the Na D fits,
which were previously poorly constrained.  Due to the low
signal-to-noise of the continuum, the doublet ratio and covering
factor are uncertain.  The results of the fitting are presented in
Table 3.    

Without the addition of \ion{K}{1}, the Na D profile for NGC 1569 is
blended with Galactic absorption and atmospheric emission.  The stellar
profile from the starburst is constrained by \ion{Mg}{1} absorption,
leaving a range of possibilities for the central velocity of the
interstellar Na D.  We find \ion{K}{1} absorption at -24 km s$^{-1}$,
thereby defining the velocity at which the interstellar Na D lies.
Due to the Fraunhofer absorption, the normalization of the continuum
in the NGC 1569 spectrum is likely imperfect and the column densities
in Table 3 are highly uncertain.  The \ion{K}{1} spectra do not
provide a useful column density constraint. 


\section{Dynamics of the Cold Neutral Gas}

Although we sample just one sightline, we can estimate how much of the
CNM is brought out into the outflow, if any.  Only five of the sample
galaxies show interstellar sodium absorption, just three of which are
dwarfs.  Due to the intricacies in geometry and modeling of the
individual galaxies, as well as the small number of sample galaxies,
it is preferable to look at each galaxy individually. Below we discuss
the galaxies independently, detailing the information gleaned from
their spectra, and presenting a model of the starburst kinematics and
any outflow/infall structure. Ideally, we would like to be able to
calculate the amount of cold, neutral gas flowing into the IGM, when
possible.  This requires knowledge not only of the kinematics and
column densities, but also the geometry of the starbursting region
itself.    


\subsection{Calculation of Superbubble Masses}

One important application of measuring the column density of a
superbubble or shell region around a starburst is to measure the mass
of gas being expelled from the starburst.
We can do this, so long as we make a few assumptions.  We use the
wind/bubble model introduced by Weaver et al. (1977), wherein a
starburst is surrounded by concentric shells of (in order of
increasing radius) a free-flowing wind, a shocked wind, and a shocked
shell of neutral ISM. The Str\"omgren sphere of the stars may be
trapped in the shell, or may completely photoionize the region.
We postulate that the H$\alpha$ emission is found in the warm, dense
ionized gas region, while the Na D absorption is found in the colder,
shocked neutral gas region.  This model is applicable in light of the
expanding shells of warm gas seen in H$\alpha$ images by Martin 1998.
We assume the shells are perfectly spherical, making the mass
a function of the column density within the shell and the area of the
shell. Given a shell of radius R, the mass of the shell in neutral
hydrogen is  
\begin{equation}
\mbox{M(\ion{H}{1})} = 4\pi R^2 \mbox{N(\ion{H}{1})}\mu m_H
C_f^{-1}.   
\end{equation}
All of the following models use parameters (R, Z/Z$_\odot$) as are
detailed in Table 4 and \ion{Na}{1} column densities as given in Table
3 and transformed into N(\ion{H}{1}) using Equation 4.  The column
densities assume ionization fractions and depletion parameters as
discussed in \S3.3, namely that $f_Df_{ion} = 100$. 


\subsection{Individual Galaxies}

\subsubsection{NGC 1569}

The sightline to NGC1569 intersects the midplane of the disk.  We
observe the shell 1569-C in H$\alpha$ at a radius of $\sim$3'' (32 pc)
from the continuum source.  This is most likely the same region in
which M\"uhle et al. (2001) parameterized an \ion{H}{1} ``hole,''
associated with the starburst, 1 kpc in diameter expanding at 7 km
s$^{-1}$.  The neutral gas outflow we observe in the nearby dwarf
galaxy NGC 1569 is moving more slowly than the bipolar outflow seen in
H$\alpha$ by Martin (1998).  If we assume the neutral sodium is
spatially coincident with the edge of this \ion{H}{1} hole, i.e. the
neutral sodium is swept up in a slowly expanding shell around the
starbursting area, then we find the total mass of cold, neutral gas
swept up to be $2.2\times 10^7$ M$_\odot(R/100 \mbox{ pc})^2$. The
\ion{H}{1} mass of this galaxy is found from radio observations to be
$8.4\times 10^7$ M$_\odot$. Therefore, the neutral gas in the shell
modeled is a small fraction ($\lesssim$10\%) of the total neutral gas
mass of the galaxy.    

Martin et al. (2002) imaged NGC 1569 in X-rays with {\it Chandra} and
argued that the observed outflowing wind is strong enough to blow
through the galactic halo. In fact, they found that nearly all of the
metals ejected by the starburst are carried into the IGM by the
galactic superwind, and are primarily from the stellar ejecta itself,
not the entrained ISM.  Therefore, it is not surprising that this
galaxy shows the smallest amount of outflowing neutral gas of our
entire sample. The outflowing mass is measured to be $1.4 \times 10^6$
M$_\odot$ in hot gas and $8.8 \times 10^6$ M$_\odot$ in warm gas
(Martin 1998).  In contrast, we find $2.2 \times 10^7$ M$_\odot$ in
\ion{H}{1} for a 500 pc (radius) shell. 

From Mg b band absorption we know there must be a stellar component of
the Na D profile at the systemic velocity.  By fitting a corresponding
stellar component to the Na D profile, and by fitting the atmospheric
absorption and Galactic emission, we find there is one interstellar
absorption component required for a good fit (see Figure 5).  This
gas, blueshifted 24 km s$^{-1}$ from the systemic, corresponds to the
\ion{K}{1} absorption seen at the same velocity. We use a value of
v$_{sys}$ = -40 km s$^{-1} \pm$ 0.6 km s$^{-1}$, obtained from the Mg
b band absorption in the stellar atmospheres within the starburst.  It
is important to note that our value of the systemic velocity is
somewhat lower than the typical values for this  galaxy:  Meier et
al. (2001) used $^{12}$CO to find a systemic velocity of -68 km
s$^{-1}$, while Stil \& Israel (2002) used the galaxy's \ion{H}{1}
rotation curve to find a mean systemic velocity of -82 km s$^{-1}$. In
fact, when comparing the heliocentric velocity of the Na D with the
above values of the systemic velocity, it is possible that while the
Na D is blueshifted in relation to the \ion{Mg}{1}, it may be at the
systemic velocity or even redshifted.  Additional uncertainty in the
analysis of the cold interstellar gas arises from possible absorption
by neutral gas in intervening Galactic intermediate- or high-velocity
clouds (Kuntz \& Danly 1996).  Moreover, \ion{H}{1} in the Galactic
thin disk is seen towards NGC 1569 (Hartmann \& Burton 1997), and
could overlap in velocity with the lines from NGC 1569.  


\subsubsection{NGC 1614}
We observed NGC 1614 in order to extend our observations beyond the
scale of dwarf galaxies and into a higher luminosity class.  This
relatively close LIRG (D = 64 Mpc for H$_o$ = 75 km s$^{-1}$
Mpc$^{-1}$) has a violent starburst in the central region which is
thought to have been triggered by interaction with another massive
galaxy, as evidenced by outer structures including plumes, tails,
and a possible secondary nucleus as seen by Alonso-Herrero et
al. (2001).  We observe the primary/nuclear starburst.  The disk of
this galaxy is inclined at $i = 51^\circ$ to the line of sight
(Alonso-Herrero et al. 2001), so our sightline intersects the
outflow. 

Our observations of this galaxy successfully resolve the absorption
profile into two separate components -- a weaker outflowing component
at -149 km s$^{-1}$, and a stronger infalling component at +70 km
s$^{-1}$.  The spectra of \ion{Na}{1}, \ion{K}{1} $\lambda 7665$, and
H$\alpha$ are shown, along with the individual components of
\ion{Na}{1} and \ion{K}{1}, in Figure 6.  Alonso-Herrero et al. (2001)
use the integrated far-infrared luminosity of the galaxy to find a
star formation rate of 52 M$_\odot$ yr$^{-1}$.  Combined with their
value of the total gas mass fueling the starburst ($3\times 10^8$
M$_\odot$), we find the starburst age to be 5.8 Myr.    Using this age
and the velocity of the outflowing neutral sodium absorption
component, we can determine the radius of the shell assuming a
constant velocity expansion.  We find in this model that the
outflowing component supports a shell 1 kpc in radius, with a total
outflow mass of $2.1\times 10^9$ M$_\odot (R/1 \mbox{ kpc})^2 $.
Assuming the redshifted/infalling component is a spherical shell, we
parameterize it in terms of a bubble with radius 1 kpc. This case
gives a total infalling mass of $2.9\times 10^9$ M$_\odot (R/1 \mbox{
  kpc})^2$.   The total \ion{H}{1} mass of this galaxy is given by
Bushouse (1987) to be $1.7 \times 10^9$ M$_\odot$, so the simple shell
geometry cannot completely model the cold ISM of this galaxy. 

This complex system is difficult to sort out.  The completely
uncoupled kinematics of the H$\alpha$ and Na D spectra suggests that
the shock fronts and ionization fronts are moving independently.  It
is possible that we are seeing redshifted absorption due 
to neutral gas in one of the filaments of gas or spiral arms of the
galaxy -- the violent interaction of NGC 1614 colliding with another
galaxy has created a complex system of filaments and hot spots which
may cause the infall of cold, neutral gas.  Alternatively, there may
be separate continuum sources, one redshifted and one blueshifted, and
at least one of which is offset from the galactic rotation axis. 


\subsubsection{NGC 2363}
NGC 2363 is an extremely luminous giant \ion{H}{2} region connected to
NGC 2366.  Our observations of the nuclear starburst show a very
straightforward double-peaked emission line structure with components
at an average velocity of 83 km s$^{-1}$ and 107 km s$^{-1}$. This is
in agreement with the models of Roy et al. (1991, and references
therein), who observed this galaxy in $[$O {\sc iii}$]$
($\lambda$5007) and also found very clear line splitting.  We see no
Na D absorption, and place an upper limit on the column density of
N(Na I) $\leq 6\times 10^{11}$ cm$^{-2}$.  

With our emission line data -- both spectra and echellegrams -- we can 
support the claim of Roy et al. (1991) that there is a relatively
simple system with one superbubble (presumably expanding) around this
giant \ion{H}{2} region.  Our $[$O {\sc iii}$]$ line splitting of 29.4 km
s$^{-1}$ (i.e. an expansion velocity of 14.7 km s$^{-1}$), compared
with the data of Roy et al. (1991), leads us to believe that we are
looking at a region about 7 arcsec -- or a physical size of 122 pc at
a given distance of 3.6 Mpc -- from the center of the
superbubble. From this distance and the blending of the emission lines
in our data, we can deduce that we are seeing the edge of the bubble.
The difference in the size of the bubble given by our measurements and
those of Roy et al. (1991) can be attributed to the difference in
assumed distance to the galaxy.  Our upper limit on the total mass of
cold neutral gas is $1.2\times 10^6$ M$_\odot (R/122\mbox{ pc})^2$.
The ionized mass of NGC 2363 was found by Luridiana, Peimbert, \&
Leitherer (1999) to be $3.4\times 10^6$ M$_\odot$, which indicates
that the outflowing cold gas is a small fraction of the galactic mass.

The absence of Na D absorption tells us that the ionization front
created by the starburst may have ``caught up'' to the shock front, so
that the ionization front is no longer trapped within the expanding
shell. Hence, we find a lack of neutral gas, even though there is very
obviously a shell present both in previous observations and in our
H$\alpha$ observations. This results in a ``leaky,'' expanding
\ion{H}{2} region, with ionizing radiation escaping beyond the edge of
the shell.    


\subsubsection{NGC 4214}
This nearby dwarf galaxy went through a starburst phase
which started roughly 10$^7$ years ago (Huchra et al. 1983).  We
observed two different positions in the galaxy, as discussed in \S2.
NGC 4214-1 (north-west complex, J2000 R.A.=12\fh15\fm39\fs48,
Decl.=36\arcdeg19\arcmin31\farcs00) corresponds to a bright super
star cluster (SSC) and likely the galactic center, while NGC 4214-2
(south-east complex, R.A.=12\fh15\fm37\fs99,
Decl.=36\arcdeg19\arcmin44\farcs20) is made of several smaller
star-forming knots. We will discuss these regions separately due to
their disparate geometry and kinematics.  Note that our positions NGC
4214-1 and -2 correspond to knots 1 and 2 of Sargent \& Fillipenko
(1991) and regions {\sc i} and {\sc ii} in MacKenty et al. (2000). 

\subsubsubsection{\it NGC 4214-1}
NGC 4214-1 is the brightest starburst knot in this galaxy.
The central knot in this region is a massive SSC containing several
hundred O stars (Leitherer et al. 1996; Sargent \& Fillipenko 1991).
Observations of ionized and neutral hydrogen around this massive
star-forming site have found large cavities in the ISM, suggesting
material was blown out of these areas by stellar winds and supernovae.
The starburst can be characterized, using the stellar population, as
an instantaneous starburst aged around 3.5 Myr (Cervi\~{n}o \&
Mas-Hesse 1994).   

No \ion{Mg}{1} absorption is seen in this spectra as there is low
signal-to-noise near the Mg b triplet.  Additionally, the Na D
absorption seen is at the local systemic velocity, so we cannot make a
conclusive statement about the nature of the absorption.  However,
accounting for the upper limit on both the equivalent width of Na D
(0.04 \AA) and Mg b (0.03 \AA), we can say that upon comparison with
equation (1) and Figure 3, it is highly likely that the Na D
absorption seen is stellar.  If this is not the case and the
absorption is interstellar, then this may be an extension of the
observations of Leitherer et al. (1996), who observe interstellar UV
lines at the systemic velocity.   Assuming the Na D absorption is
completely interstellar (i.e. an upper limit), then the covering
factor is 0.52 and the column density of \ion{Na}{1} is $5.87\times
10^{11}$ cm$^{-2}$. Assuming a bubble radius of 700 pc (an educated
guess based on the H$\alpha$ morphology seen in Martin 1998), we find
an upper limit on the mass of the stationary shell of $2.5\times 10^7$
M$_\odot (R/700 \mbox{ pc})^2$. 

\subsubsubsection{\it NGC 4214-2}
This region has a much different morphology than NGC 4214-1, with
ionization by three small OB associations ($\sim$4,000 roughly coeval
OB stars observed as H$\alpha$ knots). Our H$\alpha$ profile is
double-peaked, showing a bright emission line blueshifted from
v$_{sys}$ and a weaker redshifted component, suggesting an expanding
shell in H$\alpha$. Indeed, in looking at the H$\alpha$ echellegram,
one can easily see the ``hole'' which indicates the center  of the
shell (see Figure 7). The Na D and H$\alpha$ spectra are plotted
against velocity for NGC 4214-2 in Figure 8. Since we have no blue
spectrum for this position, we cannot use \ion{Mg}{1} to define the
nature of the gas.  However, the Na D profile is blueshifted from the
local systemic velocity as given by CO (Becker et al. 1995); this
leads us to believe the absorption is interstellar.   

The neutral sodium is observed to be expanding at 23 km s$^{-1}$, with
a width of 40 km s$^{-1}$.  The H$\alpha$ and Na D profiles have
the same Doppler width.  The kinematics indicate that there is an
expanding neutral shell and an ionized shell.  The ionized shell is
presumably interior to a concentric shell of neutral gas, and the
H$\alpha$ shell seems to be expanding a few km s$^{-1}$ faster than
the Na D shell (i.e. is more blueshifted).  
Assuming a single, expanding shell, the total mass of neutral gas
flowing out of NGC 4214-2 is $\sim7.5 \times 10^7$ M$_\odot (R/500
\mbox{ pc})^2$. 

NGC 4214 has a total \ion{H}{1} mass of $1.9\times 10^9$ M$_\odot$
(Kobulnicky \& Skillman 1996; McIntyre 1998).  The combined mass of
the maximum possible outflow in NGC 4214-1 and the outflow in NGC 4214-2
is $\sim10^8$ M$_\odot$. This indicates that while the
outflow is significant, it is but a fraction of the total neutral
ISM; its gas structure as a whole should be considerably affected by
the starbursting region.  


\subsubsection{NGC 4449}
This nearby galaxy has an unusually large B-band luminosity --
brighter than any other dwarf in this sample (see Table 1).  
{\it Chandra} X-ray imaging shows an expanding superbubble
within a cavity of H$\alpha$ emission in the region we observe
(Summers et al. 2003), and \ion{H}{1} observations show an unusually
massive galactic halo (Bajaja, Huchtmeier, \& Klein 1994).  Imaging in
H$\alpha$ (Martin 1998) shows a system of large shells breaking out
along a nuclear bar structure, with smaller bubbles growing along the
edge of the galactic disk.  It is possible that these smaller bubbles
merge into a larger bubble.    

The HIRES spectra show that the cold, neutral gas in NGC 4449 is
kinematically uncoupled from the warm, ionized gas as seen in
H$\alpha$ (see Figure 9).   The region we observe is the
center of the H$\alpha$ emission (Hunter et al. 1998), and the neutral
sodium we see expanding at 34 km s$^{-1}$ is most likely a product of
several superbubbles expanding out of the galactic plane along the
line of sight.  Indeed, the column density, N$_{NaD}$ = $5.1 \times
10^{13}$ $(f_Df_{ion}/100)$ cm$^{-2}$, is larger than in any other
dwarf.  The unusually  large sodium column may be due to additional
neutral gas in the halo, explaining the existence of as much as 300
times more neutral matter in this galaxy as compared to NGC 1569 or
NGC 4214 (Bajaja et al. 1994).  If the CNM being expelled from the
nuclear region is being pushed by the conglomerate effect of the
H$\alpha$ bubbles, we can approximate the geometry of the neutral gas
by a single bubble with a radius $\sim$1 kpc.  This radius is on the
order of the radius of the largest complex seen in H$\alpha$ imaging,
which Martin (1998) calls NGC 4449-A (radius = 950 pc).  Given the
high column and radius, we find that this shell contains a total
neutral gas mass of $7\times 10^9$ M$_\odot (R/1 \mbox{ kpc})^2$.
From radio observations, Bajaja et al. (1994) find a total galactic
\ion{H}{1} mass of $2.3\times 10^9$ M$_\odot$, which suggests that we
may be seeing the galaxy's massive halo along our line of sight. 


\subsubsection{NGC 5253}

We detect Na D absorption at a 3$\sigma$ level in NGC 5253, but as
discussed in \S3.1, this absorption is believed to be stellar because
it is at the same velocity as the \ion{Mg}{1} absorption and it falls
directly on the relation for stars (see Figure 3).  Marlowe et
al. (1995) observe two kpc-scale superbubbles in H$\alpha$ flanking
the galactic center along its minor axis.  We see these in emission
only: a brighter line blueshifted by 39 km s$^{-1}$ and a weaker line
blueshifted by 73 km s$^{-1}$.  Our sightline to the nuclear starburst
(5253-1; Calzetti et al. 1997) could intersect either bubble, or both;
the geometry is unclear.  The upper limit on interstellar sodium
absorption we detect is N(NaD) = $2\times 10^{11}$ cm$^{-2}$, which
gives a mass of $\sim 10^7$ M$_\odot$ (R/870 pc)$^2$ using the
bubble radius seen in H$\alpha$ by Martin (1998). This is still a
large amount of mass, so this is by no means a definitive statement
that there is no expanding shell; our sightline simply misses the
shells as shown in Martin 1998 (Figure 2$j$ in that paper).  The
\ion{H}{1} mass of NGC 5253 is roughly $1.4\times 10^8$ M$_\odot$
(Kobulnicky \& Skillman 1995), indicating that the mass of cold
outflowing gas detected is small compared to the total neutral gas
mass of the galaxy.     


\subsubsection{M82}

Our observations of Na D in M82 show an extremely complex absorption
system, best fit with five separate absorption doublets.   The two
blueshifted components are the strongest; there is a single component
within 4 km s$^{-1}$ of the systemic velocity (v$_{sys}$ = 214 km
s$^{-1}$ -- see Table 1), and there are two fairly strong redshifted
doublet lines.  All of the components are highly saturated with the
exception of the redmost component, which is optically thin (i.e. the
doublet ratio = 2).     

M82 is viewed edge-on, and our sightline intersects the nuclear region
of the galactic disk.   A bright, bipolar outflow is observed out to
distances of $>$11 kpc along the minor axis of the galaxy (Lehnert,
Heckman, \& Weaver 1999). CO observations by Wei\ss~et al. (1999),
combined with the schematic diagram presented by Ohyama et al. (2002,
see their Figure 7), leads us to believe we are looking through
several filaments separated by regions of warmer, nuclear
emission. The CO images are centered on the most powerful supernova
remnant in the galaxy, SNR 41.9+58, roughly 17\arcsec~from our
sightline (R.A.=09\fh55\fm53\fs1,
Decl.=69\arcdeg40\arcmin47\farcs2). Using our value of the systemic
velocity (214 km s$^{-1}$, Lo et al. 1987), the most redshifted
component in our Na D spectrum is at approximately the same velocity
as the CO.  This suggests that the redshifted atomic gas is absorption
in the galactic disk close to the molecular gas, near the starburst. 

H$\alpha$ images (Martin 1998) and \ion{H}{1} images (Wills,
Pedlar, \& Muxlow 2002) show that a large, expanding superbubble is
seen along the minor axis of the galaxy, and is likely made up of
several smaller bubbles forming an outflow as they merge.  We
can make an educated guess on the distance from the continuum source
to the \ion{Na}{1} absorbing gas by using the ages derived from
H$\alpha$ shell expansion in Martin (1998), and find the dynamical age
of the shells to be $\sim$8 Myr. This gives shell sizes on the order
of a few hundred pc. Assuming solar metallicity, we can find the mass
expelled in the simple case of independently expanding spheres.  The
shells range in neutral gas mass from $\sim 5\times 10^6$ M$_\odot (R/286
\mbox{ pc})^2$ to $8\times 10^7$ M$_\odot (R/500 \mbox{ pc})^2$. These
are relatively small compared to the other similarly luminous galaxies
where \ion{Na}{1} absorption is detected; it should be noted that the
radii given, and therefore the masses, are a lower limit. 


\subsubsection{I Zw 18}

I Zw 18 is important to study due to its extremely low metallicity -
O/H $\approx$ 0.02(O/H)$_\odot$ - and high \ion{H}{1} column, making
it a prime candidate for a young galaxy undergoing its first phase of
star formation (Lequeux et al. 1979).  Moreover, the galaxy has a
relatively low escape velocity, and feedback from star formation could
drastically affect the galaxy (Martin 1996).  We observed I Zw 18
hoping to see absorption from cold gas in the largest \ion{H}{2}
region in the galaxy. However, we see no Na D absorption and our
spectra show a relatively featureless continuum, with a prominent
hydrogen Balmer series and weak emission from [\ion{O}{3}]
($\lambda\lambda$4363, 4959, 5007), and only the stronger line from
the [\ion{S}{2}] doublet at $\lambda$6716. We can place an upper limit
on the column density of Na D of $\sim$10$^{12}$ cm$^{-2}$,
using a S/N $\sim$4 from the continuum and an assumed line width of 30
km s$^{-1}$ -- on par with the line widths in other dwarfs.  This
translates to an upper limit of $2.1 \times 10^9$ M$_\odot (R/970
\mbox{ pc})^2$ on the mass of neutral gas in {\sc I} Zw 18, if the
cold, neutral gas is entrained in the 970 pc superbubble
parameterized in Martin 1998.  It would be somewhat surprising to find
even this much neutral sodium in this bubble, considering the
incredibly low metallicity of this galaxy.  Using a metallicity of 2\%
solar, we find an upper limit on the \ion{H}{1} column of $\lesssim
10^{22}$ cm$^{-2}$.  This is calculated given an abundance 
[Na/H] = -1.70, which is consistent with the value of [Fe/H] = -1.76
derived from {\em FUSE} observations by Aloisi et al. (2003).  This
roughly result agrees with previous measurements of the \ion{H}{1}
column of a few $\times$ $10^{21}$ cm$^{-2}$ (Aloisi et al. 2003;
Lecavelier des Etangs et al. 2004; van Zee et al. 1998), and is
consistent with a $\sim$2\% solar abundance of sodium.   


\section{Conclusions}

We have obtained high-resolution Echelle spectra of six dwarf
starburst galaxies, NGC 1614, and M82 in order to study the cold
interstellar gas in dwarf starbursts.  Interstellar neutral sodium
column densities were obtained by measuring the equivalent width of
the Na D absorption doublet.  We find that out of the eight galaxies,
NGC 1569, NGC 1614, NGC 4214-2, NGC 4449, and M82 unambiguously show
interstellar sodium absorption, while NGC 2363, NGC 4214-1, NGC 5253,
and I Zw 18 do not. The dwarf galaxies NGC 1569, NGC 4214, and NGC
4449 exhibit single-component outflows of neutral gas.  NGC 1614 and
M82 have multiple interstellar components, some outflowing and some
infalling, and both galaxies show more absorption from outflowing gas
than any dwarf galaxy. 

The dwarf galaxies show trends similar to brighter, larger
galaxies, but on a smaller scale.  While samples of LIRGs and ULIRGs
are shown to have average outflow speeds of $\sim$100 km s$^{-1}$
(HLSA) and $\sim$700 km s$^{-1}$ (Rupke et al. 2002; Martin \& Armus
2004), respectively, the three dwarfs show an average outflow speed of
only $\sim$30 km s$^{-1}$.   Additionally, most ULIRGs (e.g. 73\% in
Rupke et al. 2002) show an outflow, whereas only half of the galaxies
in our sample have an outflow region.  Spectral lines in dwarf
galaxies also have  smaller velocity spreads.  

It is particularly interesting to see how complex the absorption line
spectra become at high resolution.  M82 is an excellent example of an
absorption line system previously believed to be a single pair of
lines, but within each member of the doublet we find a wealth of
structure. We resolve at least five line pairs, showing the incredibly
complicated nature of the CNM in this galaxy.  This is a very
intriguing result and prompts us to wonder how many of the galaxies in
previous samples could be resolved into complex systems of multiple
components.  NGC 1614 shows complex absorption as well, though it has
a far smoother profile in Na D (and \ion{K}{1}) than M82. The three
dwarfs with outflows do not seem to exhibit this behavior, showing far
simpler spectra.     

For the first time we are able to combine measurements of the
kinematics of the warm, ionized gas in emission with absorption
spectra of the cold, neutral gas.  For sightlines intersecting a
single shell, we can determine whether the ionization front is trapped
inside the shock front.  This works extremely well for NGC 2363, which
is a simple expanding bubble with an ionization front that has
expanded beyond the shock front.  NGC 4214-2 also presents a
relatively simple shell, and this galaxy also shows Na D absorption.
From the kinematics we postulate that the ionization front is
expanding faster than the shock front, though it has clearly not
``caught up'' yet, as there is still a concentric, outer expanding
bubble of cold, neutral gas.  In other words, the ionization front is
still trapped in the shell.  The kinematics of Na D and H$\alpha$ are
very different in other galaxies, and cannot be easily explained by
this straightforward picture.

Combining our measurements of \ion{Na}{1} column density with previous
observations of superbubbles and supershells in the sample galaxies,
we have parameterized the mass of neutral gas (or a limiting case
thereof) flowing out of the starbursting region.  Using a simple
spherical shell model, we find the total neutral gas outflow masses to
be roughly 10$^6$ to 10$^{10}$ M$_\odot$.  NGC 4449 shows far more
outflowing cold gas than any other dwarf galaxy.  Compared to the
measurements of outflowing warm and hot gas from these galaxies
(Martin 1998), it is likely that the bulk of the energy in the outflow
is carried by the warm and hot gas, rather than cold gas.


\acknowledgements

Financial support was provided by the David and Lucille Packard
Foundation and the Alfred P. Sloan Foundation.  This research has made
use of the NASA/IPAC Extragalactic Database (NED) which is operated by
the Jet Propulsion Laboratory, California Institute of Technology,
under contract with the National Aeronautics and Space
Administration. This research has made use of the NASA Astrophysics
Data System abstract service.  The authors wish to recognize and
acknowledge the very significant cultural role and reverence that the
summit of Mauna Kea has always had within the indigenous Hawaiian
community.  We are most fortunate to have the opportunity to conduct
observations from this mountain.  We appreciate comments and
clarifications from an anonymous referee.  




\begin{figure}
\plotone{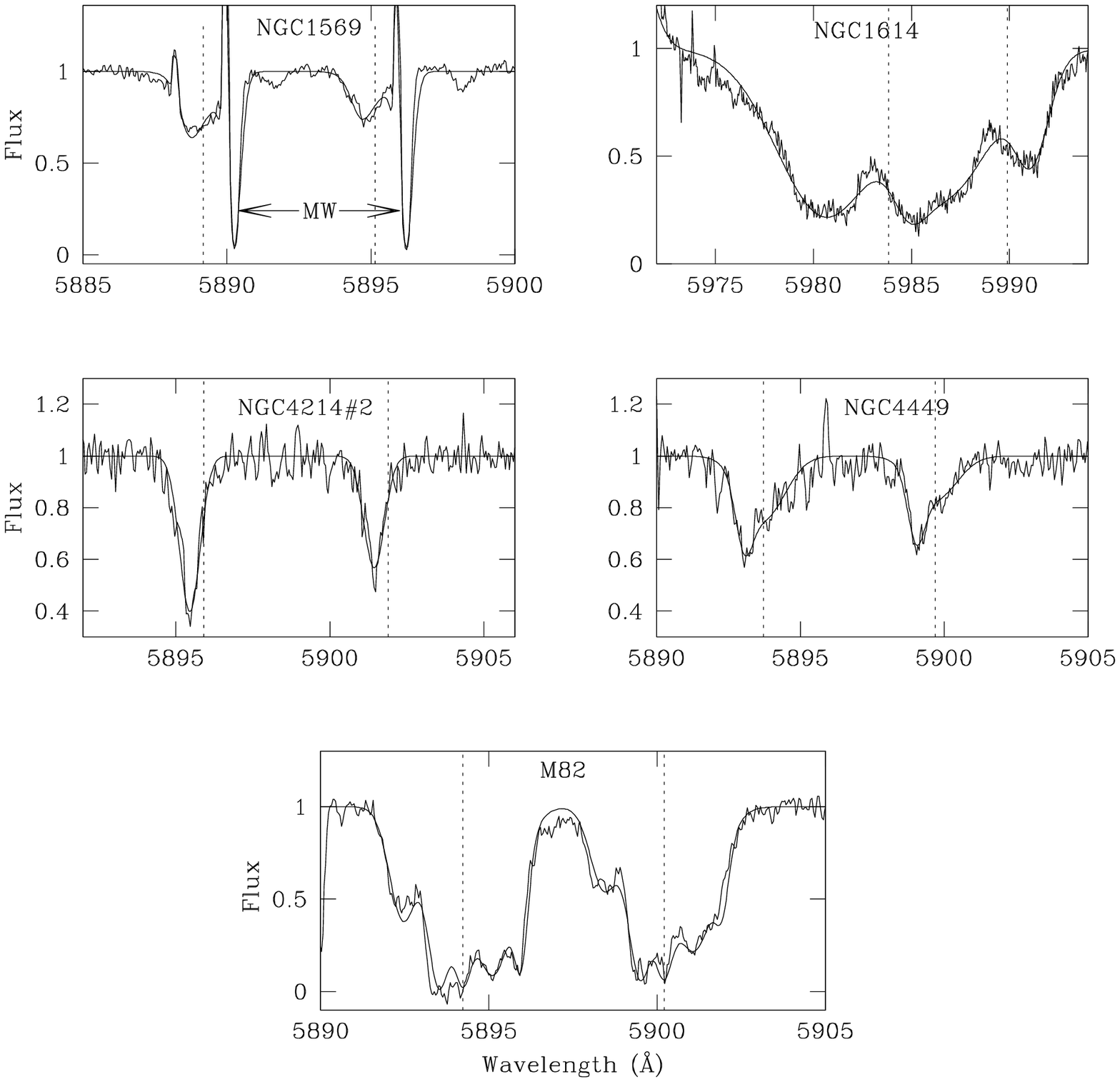}
\caption{Normalized spectra of the Na D absorption line profile in the
five galaxies in our sample where sodium was detected.  The solid line
plotted over the data is the fit from the SPECFIT program.  The dotted
lines represent the wavelengths corresponding to the systemic velocity
for each of the Na D lines, which is the velocity given by the stellar
\ion{Mg}{1} absorption, \ion{H}{1} rotation curves, or CO emission.
The narrow emission and absorption in NGC 1569 are due to the
atmosphere and the Galaxy, respectively, and are shown in detail in
Figure 5.}  
\end{figure}



\begin{figure} 
\plotone{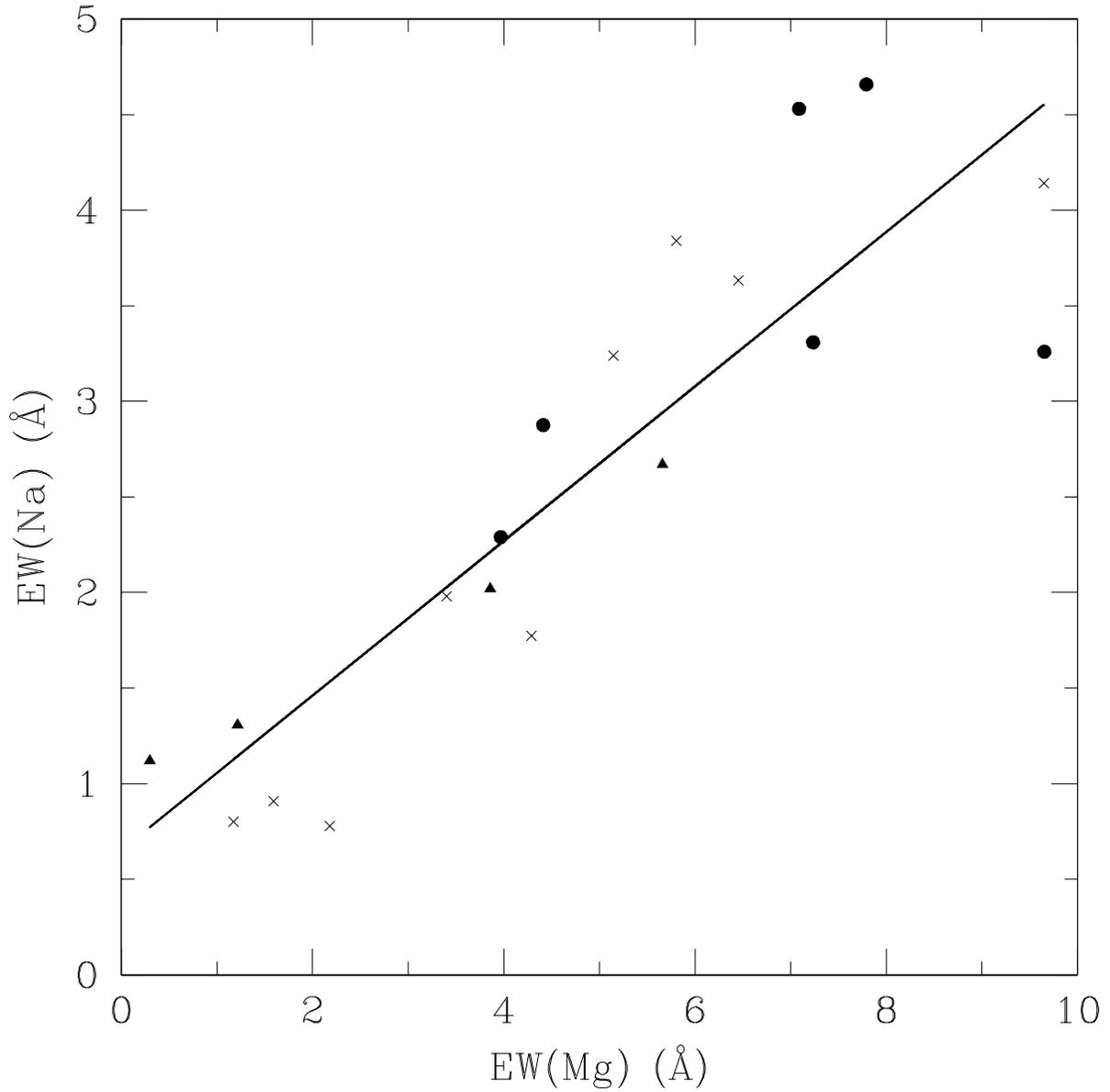}
\caption{Total equivalent width of the \ion{Mg}{1}
  absorption triplet versus that of the \ion{Na}{1} doublet using data
  from Jacoby, Hunter, \& Christian (1984).  The relevant stars for
  this application are cooler ones in luminosity classes III (crosses), II
  (filled triangles), and I (filled circles).  The solid line is a
  linear least-squares fit to the data.  This gives us a slope of
  0.40 and a y-intercept of 0.65.} 
\end{figure}



\begin{figure}
\plotone{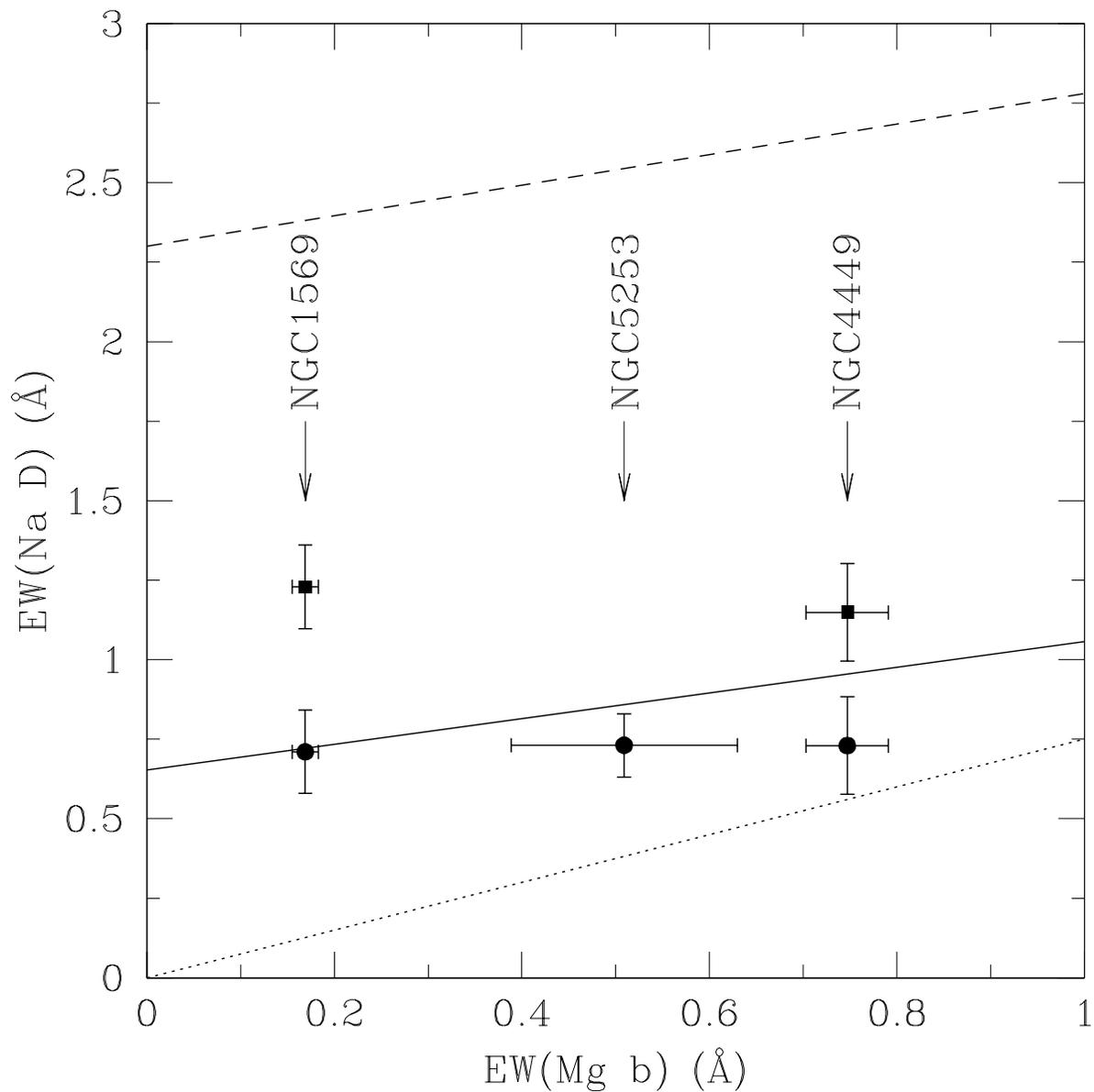}
\caption{Total equivalent width of the \ion{Mg}{1} absorption triplet
  versus stellar \ion{Na}{1} (filled circles), and total (stellar +
  interstellar) \ion{Na}{1} (filled squares), for the three galaxies
  showing \ion{Mg}{1} absorption.  Only one point is shown for NGC 5253
  because it has a single stellar component.  The dotted line is the
  model used by HLSA, the dashed line is the fit given for elliptical
  galaxies by Bica et al. (1991), and the solid line is our fit to the
  stellar spectra of Jacoby et al. (1984).}
\end{figure}



\begin{figure}
\plotone{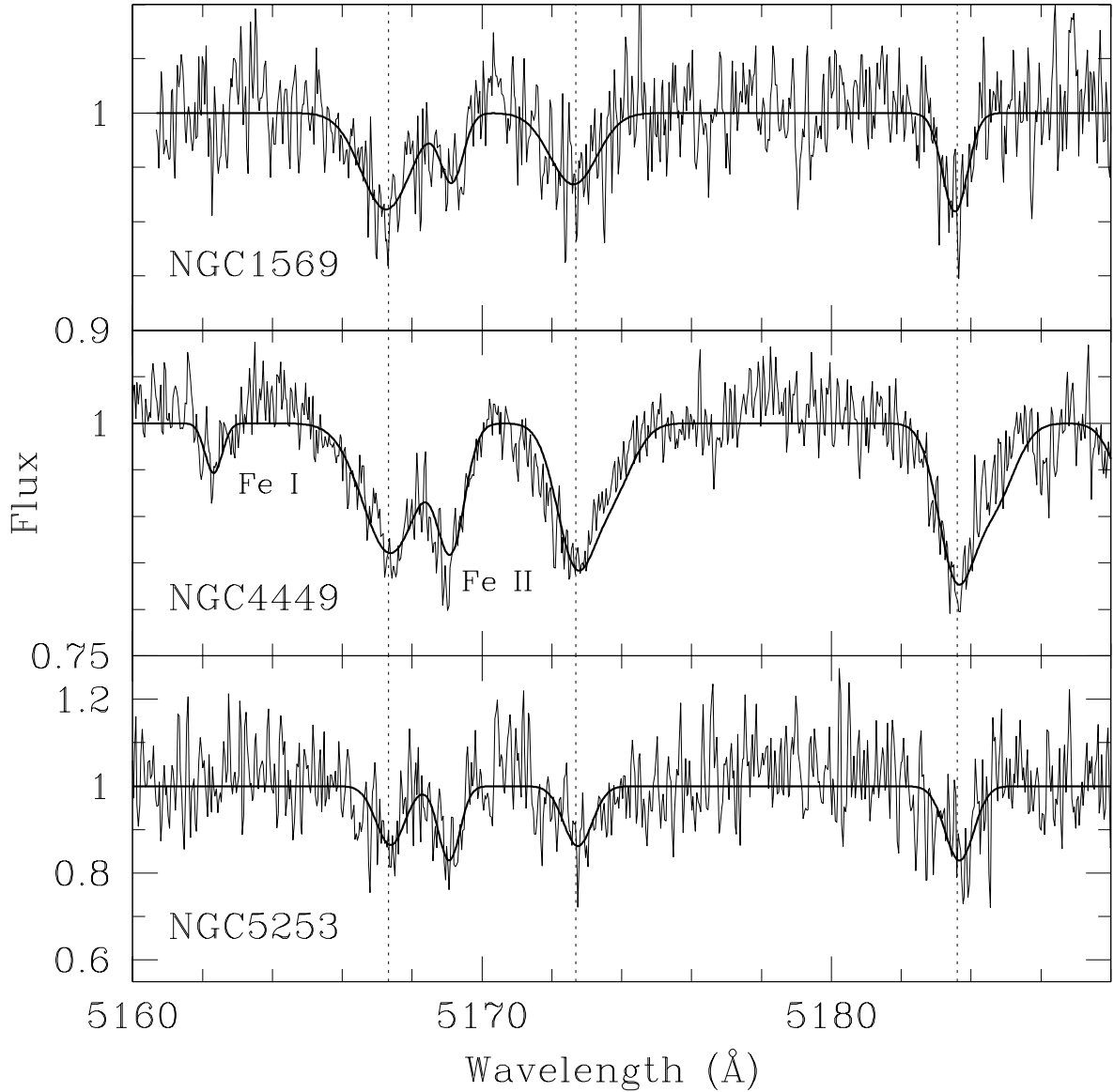}
\caption{\ion{Mg}{1} spectra for NGC 1569, NGC 4449, and NGC 5253,
  corrected for the systemic velocity by dividing by (1+v$_{sys}$/c).
  The vertical lines indicate the rest wavelengths of the Mg b band
  triplet lines. The other stellar absorption lines present are
  \ion{Fe}{1} (5162.3~\AA) and \ion{Fe}{2} (5169.0~\AA).}
\end{figure}



\begin{figure}
\plotone{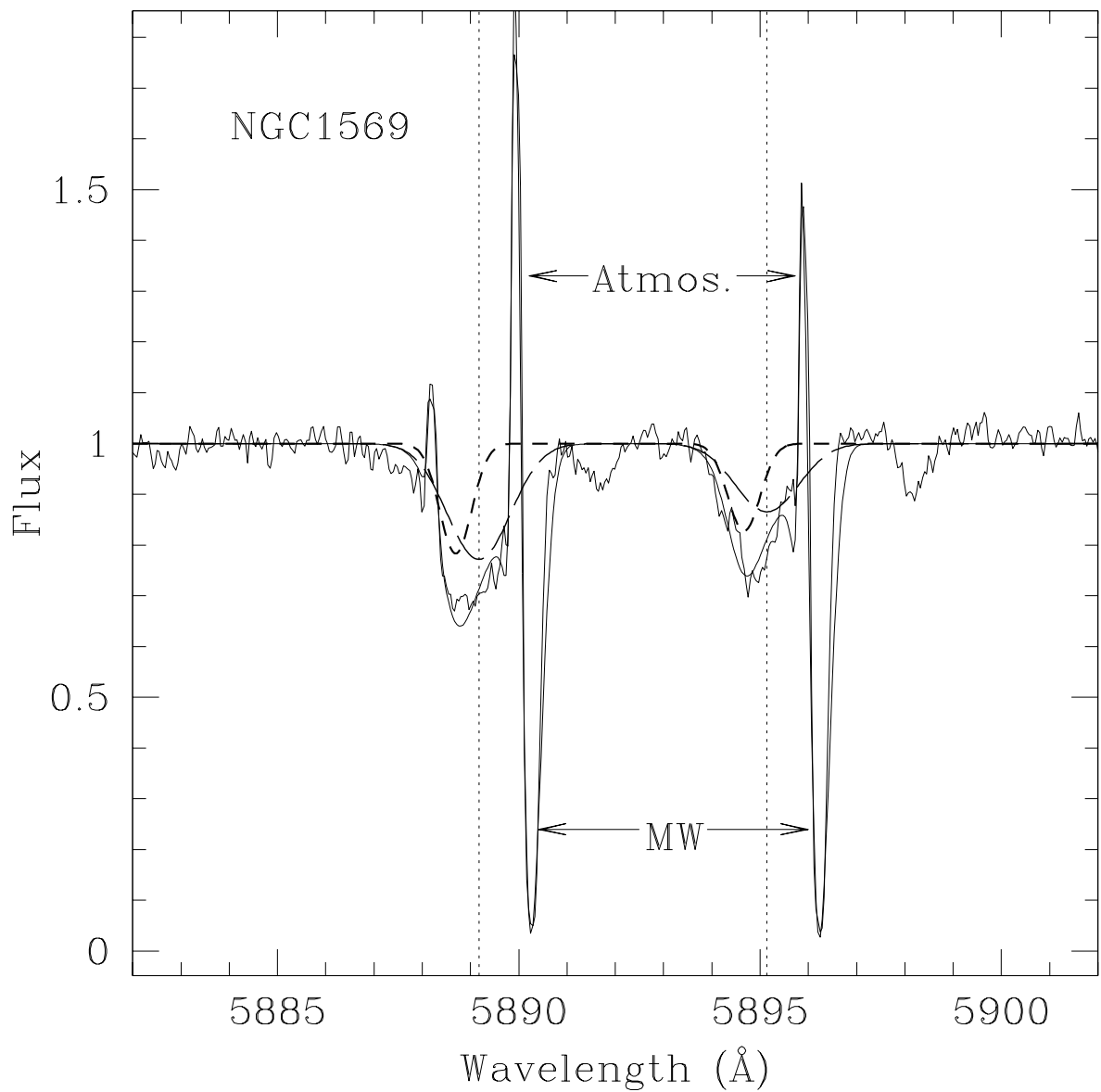}
\caption{Na D spectrum in NGC 1569, with no background subtraction.
  Vertical dotted lines are at the systemic velocity given by
  \ion{Mg}{1} absorption ($-39.7$ km s$^{-1}$).  Overplotted are the
  model obtained using SPECFIT (solid line), the components of
  interstellar sodium corresponding to the \ion{K}{1} absorption
  (short-dashed line), and the components of stellar sodium
  corresponding to the \ion{Mg}{1} absorption (long-dashed line).  The
  emission is at 0 km s$^{-1}$, due to sodium in the atmosphere, and
  is labelled ``Atmos.''  The deep, narrow absorption is due to sodium
  in the galactic plane and is labelled ``MW.''}  
\end{figure}



\begin{figure}
\plotone{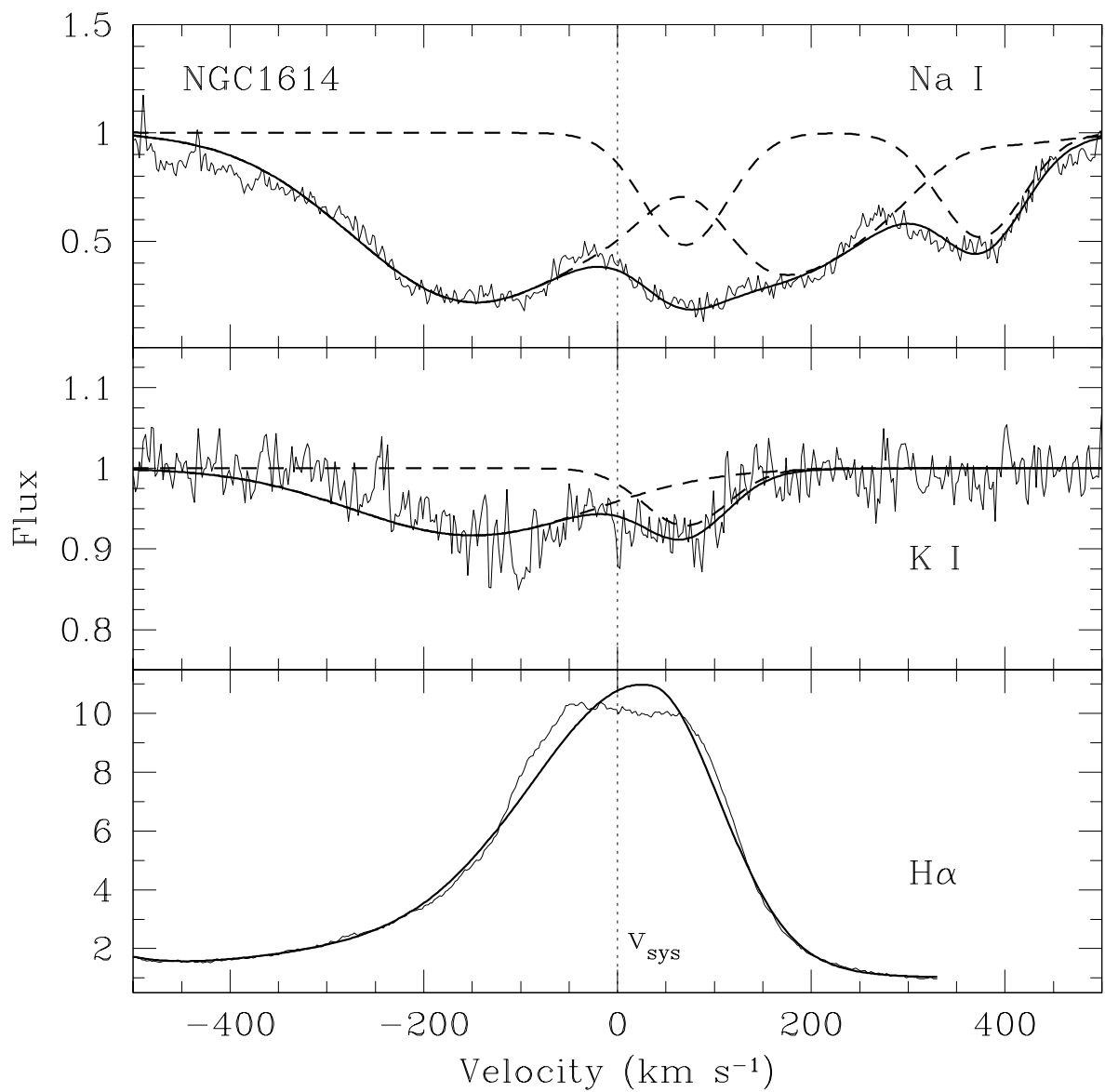}
\caption{\ion{Na}{1}, \ion{K}{1}, and H$\alpha$ spectra for NGC 1614.
  The solid lines are the Gaussian models provided by SPECFIT, the
  dashed lines in \ion{Na}{1} and \ion{K}{1} show the components.  In
  sodium, all the Na D components overlap each other, whereas
  \ion{K}{1} shows only the 7665~\AA~line.  Note the prominently
  blueshifted and redshifted components of the absorption features (at
  -149 km s$^{-1}$ and +70 km s$^{-1}$, respectively), as well as the
  complete lack of correlation with the H$\alpha$ emission. The dotted
  line shows the systemic velocity, as determined from CO observations
  by Casoli et al. (1991).  The \ion{Na}{1} was converted to a
  velocity scale using the wavelength of the blue member of the Na D
  doublet.} 
\end{figure}



\begin{figure}
\plotone{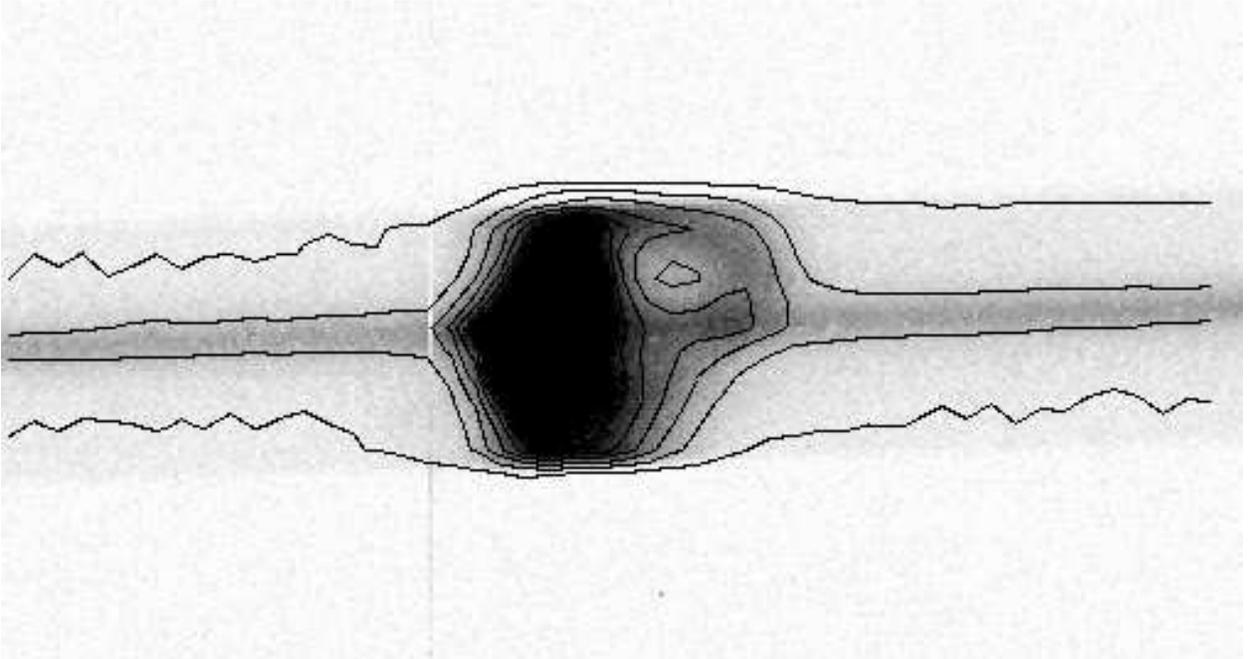}
\caption{H$\alpha$ echellegram for NGC 4214-2.  Wavelength increases
  to the right.  The vertical axis is spatially along the slit; the
  aperture is 10'' long (175 pc at a distance of 3.6 Mpc).  The total
  width of the H$\alpha$ feature is 5.7 \AA, or 260 km s$^{-1}$.  The
  ``hole'' seen on the red side of the spectral feature measures
  approximately 2'' (35 pc) by 28 km s$^{-1}$.  The contour spacing is
  linear.}
\end{figure}



\begin{figure}
\plotone{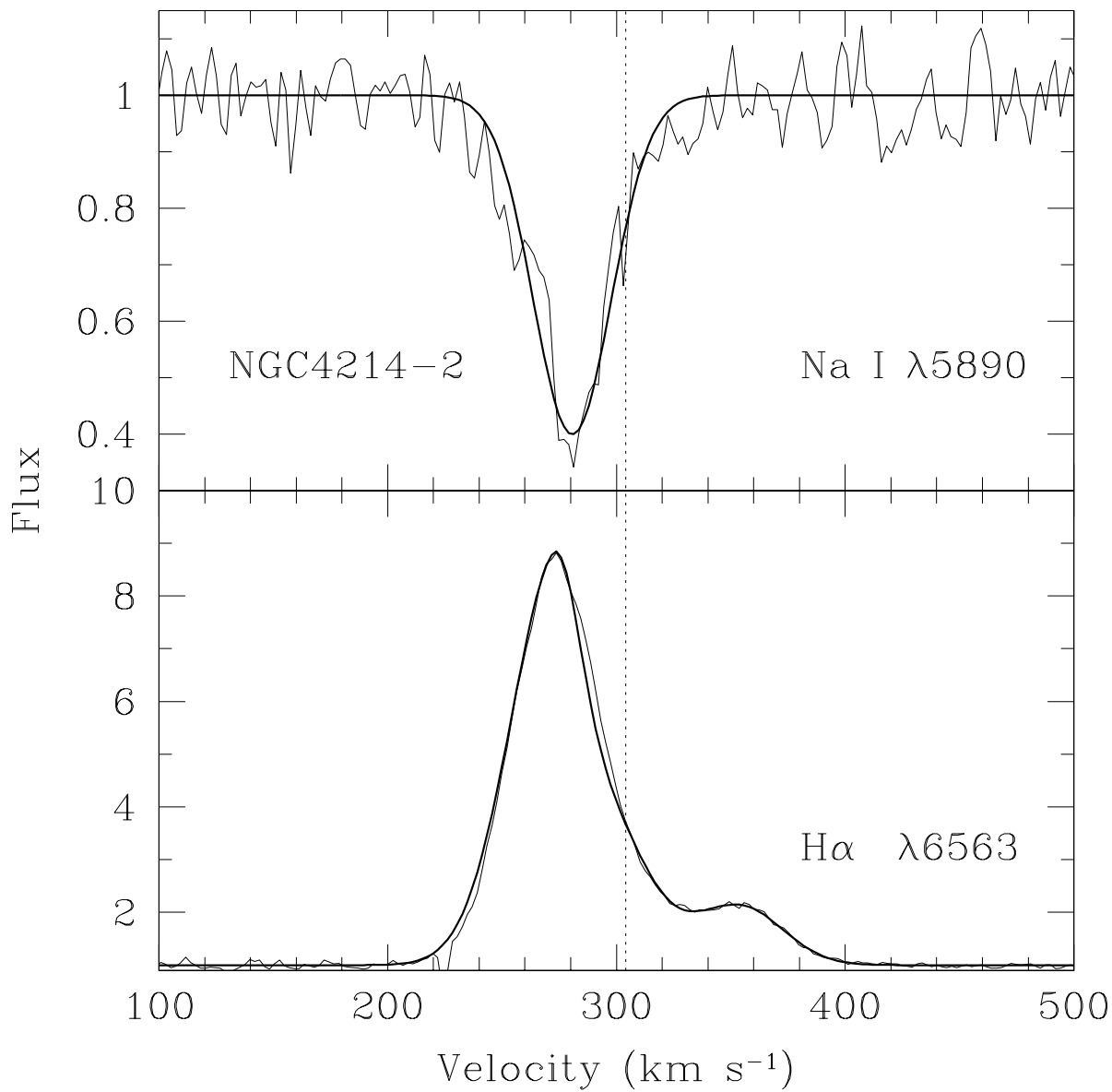}
\caption{Normalized Na D and H$\alpha$ spectra in velocity space for
  NGC 4214-2.  The spectral fit plotted over the data with a solid
  line.  The dotted line is the systemic velocity (304 km s$^{-1}$) as
  given by $^{12}$CO (Becker et al. 1995).  The top panel shows the
  blue (stronger) line of the Na D absorption doublet ($\lambda$5890),
  which is blueshifted by $\sim$24 km s$^{-1}$. The bottom panel shows
  the H$\alpha$ emission line, which is bimodal with one redshifted
  component and a brighter component blueshifted by $\sim$30 km
  s$^{-1}$.  The difference in velocity suggests that the ionization
  front is expanding faster than the shock front, although there is
  still a region of cold, neutral gas between the two fronts.} 
\end{figure}



\begin{figure}
\plotone{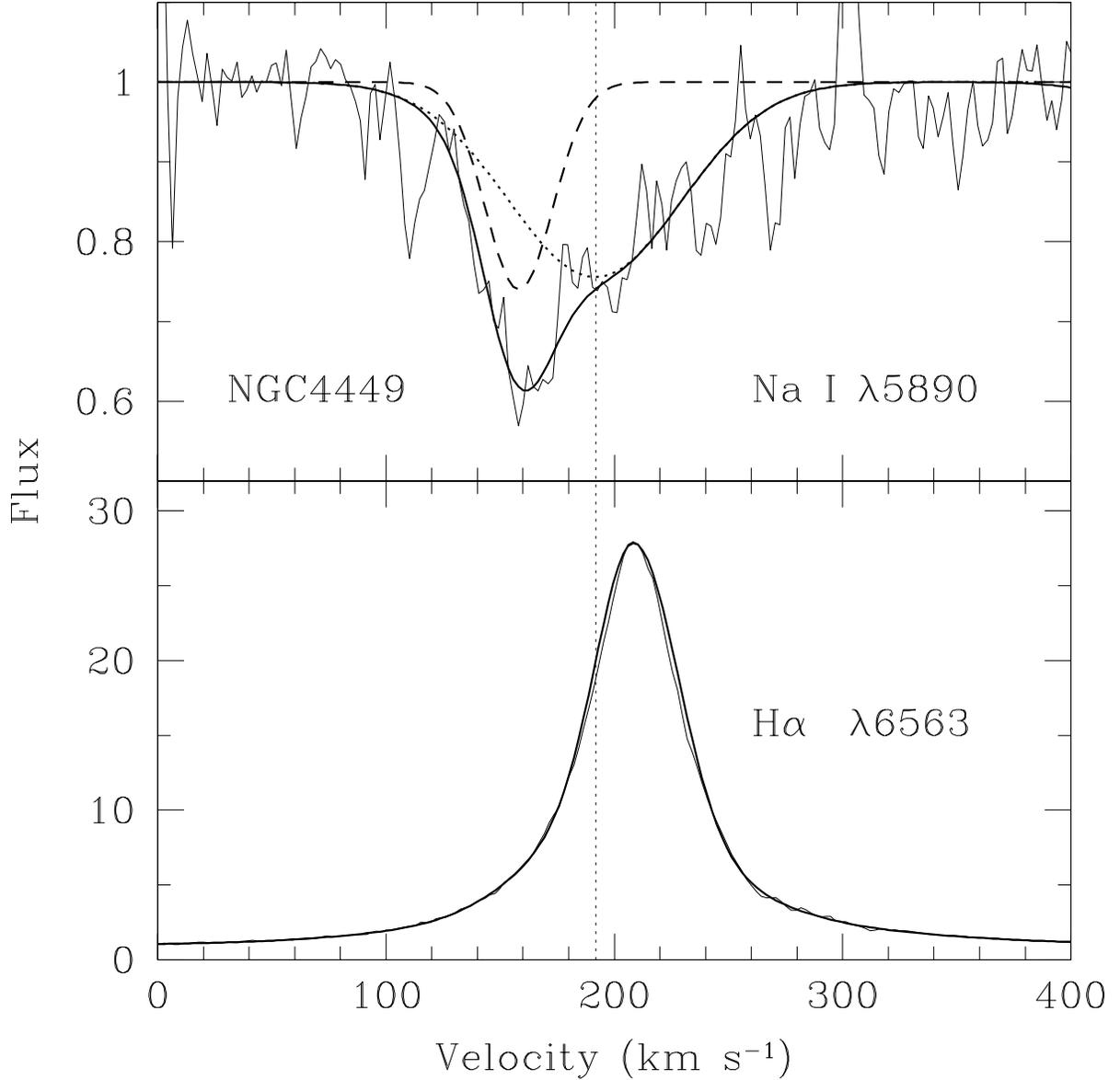}
\caption{The normalized spectra in velocity space for
  NGC 4449 with spectral fit plotted over with a solid line.  The dotted
  line is the systemic velocity as given by Mg b band absorption in
  this paper (192 km s$^{-1}$).  The top panel shows the blue
  (stronger) line of the Na D absorption doublet ($\lambda$5890),
  which is blueshifted by $\sim$34 km s$^{-1}$ and fit with a stellar
  component at the systemic velocity.  The bottom panel shows the
  H$\alpha$ emission line, which is fit with one component at the
  systemic velocity and a brighter component redshifted by $\sim$17 km
  s$^{-1}$.}
\end{figure}


\clearpage


\begin{deluxetable}{lcccccccc} 
\tablecolumns{9} 
\tablewidth{0pc}
\tablecaption{{\sc The Sample of Galaxies}}
\tablehead{ 
\colhead{Galaxy} & \colhead{Type}      & \colhead{v$_{sys}$} 
                 & \colhead{v$_{rot}$\tablenotemark{a}} 
                 & \colhead{d}   & \colhead{M$_B$}
                 & \colhead{M(\ion{H}{1})}
                 & \colhead{W$_{NaD}$\tablenotemark{b}}
                 & \colhead{References}\\
                 &                     & (km s$^{-1}$) 
                 & (km s$^{-1}$)       
                 & (Mpc)            &  & \colhead{(10$^7$M$_\odot$)}
                 & (\AA)               & }
\startdata
NGC 1569  &   {\sc I}Bm    &  -40  &   28  &  2.2  & -17.26 & 5.9 & 0.92 & 1, 2 \\
NGC 1614  &   SB(s)c pec   & 4730  &  210  & 64.0  & -20.84 & 170 &
                                                                11.25 & 3, 4 \\
NGC 2363  &  {\sc I}B(s)m  &   70  &   53  &  3.6  & -16.75 &  79 &
                                                                 ...  & 5, 2 \\
NGC 4214  & {\sc I}AB(s)m  &  304  &   35  &  3.6  & -17.65 & 110 &  
                                                1.05\tablenotemark{c} & 6, 2 \\
NGC 4449  &   {\sc I}Bm    &  192  &   65  &  3.6  & -17.86 & 135 &
                                                                 0.34 & 1, 2 \\
NGC 5253  & {\sc I}m(pec?) &  389  & $<$15 &  4.1  & -17.62 &  20 &
                                                                 0.73 & 1, 2 \\
M82      &    {\sc I}0    &  214  &  135  &  3.6  & -18.95 &  88 &
                                                                 7.98 & 7, 8 \\
I Zw 18  &       ...      
                  &  761  &   50  & 10.0  & -13.84 & 2.6 &  ...  & 9, 10 \\
\enddata
\tablenotetext{a}{Rotational velocity for NGC 1614 is from global CO 115
  GHz emission-line profiles using the half-width at 20\% of the peak
  intensity and then correcting for the inclination (HLSA).  All
  others are from Martin (1998) and references therein.} 
\tablenotetext{b}{W$_{NaD}$ given is the total equivalent width of the
  entire Na D absorption feature, i.e. both doublet lines, regardless
  of the source; it is uncorrected for covering factor.} 
\tablenotetext{c}{W$_{NaD}$ given for NGC 4214 is total sodium
  equivalent width for both NGC 4214-1 and NGC 4214-2, which have values
  of 0.18~\AA~and 0.86~\AA, respectively.}
\tablecomments{Systemic velocity value in the third column is from the
  first reference in the last column.  Absolute blue magnitude M$_B$ is
  calculated from the RC3 magnitude B$_T^0$, except in the case of
  NGC 1614, which is calculated from the value given in the NASA
  Extragalactic Database (NED) and corrected for Galactic/foreground
  extinction.  \ion{H}{1} mass is from the second reference in the
  last column.}  
\tablerefs{(1) This paper.  (2) Karachentsev, Makarov, \&  Huchtmeier
  1999. (3) Casoli et al. 1991.  (4) Bushouse 1987.  (5) NED.  (6)
  Becker et al. 1995.  (7) Lo et al. 1987.  (8) Sofue 1997.  (9)
  Aloisi et al. 2003. (10) Papaderos et al. 1996. } 
\end{deluxetable}



\begin{deluxetable}{lccccccc}
\tablecolumns{8} 
\tablewidth{0pc}
\tablecaption{{\sc Interstellar \ion{Na}{1} Column Densities}}
\tablehead{ 
\colhead{Galaxy} & \colhead{v-v$_{sys}$} 
                 & \colhead{W$_{5890}$\tablenotemark{a}}
                 & \colhead{Doublet}               
                 & \colhead{FWHM}
                 & \colhead{C$_f$} 
                 & \colhead{Z/Z$_\odot$\tablenotemark{b}}
                 & \colhead{N(\ion{Na}{1})}  \\
                 & \colhead{(km s$^{-1}$)}   
                 & \colhead{(\AA)}
                 & \colhead{Ratio} 
                 & \colhead{(km s$^{-1}$)}
                 & &
                 & \colhead{(10$^{12} $cm$^{-2}$)}}

\startdata  
NGC 1569    &  -24 & 0.20 & 1.38 &  39.7 & 0.23 & 0.25 &  0.62 $\pm$ 0.22   \\
NGC 1614    & -149 & 4.85 & 1.15 & 294.7 & 0.80 & 0.70 & 42.52 $\pm$ 0.20   \\
            &  +70 & 1.13 & 1.08 & 102.7 & 0.52 & ...  & 58.93 $\pm$ 0.20   \\
NGC 4214-2  &  -23 & 0.50 & 1.39 &  39.9 & 0.71 & 0.25 &  2.2  $\pm$ 0.28   \\
NGC 4449    &  -34 & 0.17 & 1.07 &  30.9 & 0.27 & 0.25 & 50.9  $\pm$ 0.28   \\
M82         &  -91 & 0.62 & 1.36 &  49.1 & 0.69 & 1.0  &  3.0  $\pm$ 0.17   \\
            &  -35 & 1.03 & 1.11 &  47.2 & 1.00 & ...  &  1.7  $\pm$ 0.17   \\
            &   +4 & 0.77 & 1.13 &  36.7 & 1.00 & ...  &  7.7  $\pm$ 0.17   \\
            &  +45 & 1.52 & 1.19 &  78.5 & 1.00 & ...  &  9.4  $\pm$ 0.17   \\
            &  +86 & 0.46 & 1.98 &  24.6 & 1.00 & ...  &  2.4  $\pm$ 0.17   \\ 
\\
NGC 2363    &  ... & 0.11 & ...  & ...   &  ... & 0.25 & $<$0.57  \\
NGC 4214-1  &  ... & 0.04 & ...  & ...   &  ... & 0.25 & $<$0.37  \\
NGC 5253    &  ... & 0.07 & ...  & ...   &  ... & 0.37 & $<$0.20  \\
I Zw 18
            &  ... & 0.25 & ...  & ...   &  ... & 0.02 & $<$1.27  \\
\enddata 
\tablecomments{Neutral sodium column density is contingent on
  ionization parameter and fractional depletion of sodium onto 
  grains.  These quantities are inversely proportional to the column 
  density, and therefore the given values above are estimates in that
  respect.  See section 3.3.} 
\tablenotetext{a}{W$_{5890}$ refers to the model fit of the blue line,
in Angstroms, when interstellar in nature.  When no interstellar Na D
is seen, an upper limit is given.  This measurement is uncorrected for
covering factor, though a correction is included in the column density
calculations.}
\tablenotetext{b}{Metallicity used is given in the following references: 
  NGC 1569 -- Martin et al. 2002; NGC 1614 -- 
  Storchi-Bergmann, Calzetti, \& Kinney 1994; NGC 2363 -- Luridiana,
  Peimbert, \& Leitherer 1999; NGC 4214 -- Leitherer et al. 1996; NGC
  4449 -- Skillman, Kennicutt, \& Hodge 1989; NGC 5253 -- Marlowe et
  al. 1995; M82 -- Umeda et al. 2002, and references therein; I Zw 18
  -- Martin 1996.  Solar abundances are defined in Cameron 1973.} 
\end{deluxetable}  



\begin{deluxetable}{lcccccc} 
\tablecolumns{6} 
\tablewidth{0pc}
\tablecaption{{\sc Interstellar \ion{K}{1} Column Densities}}
\tablehead{ 
\colhead{Galaxy} & \colhead{v-v$_{sys}$} 
                 & \colhead{W$_{7665}$\tablenotemark{a}}
                 & \colhead{Doublet}               
                 & \colhead{FWHM}
                 & \colhead{N(\ion{K}{1})}  \\
                 & \colhead{(km s$^{-1}$)}   
                 & \colhead{(\AA)}
                 & \colhead{Ratio} 
                 & \colhead{(km s$^{-1}$)}
                 & \colhead{(10$^{12} $cm$^{-2}$)}}

\startdata  
NGC 1569    &  -24 & 0.38 & 1.38 &  39.7 & 1.9    \\
NGC 1614    & -149 & 0.68 & 1.15 & 294.7 & 6.0    \\
            &  +70 & 0.20 & 1.08 & 102.7 & 8.2    \\
\enddata 
\tablecomments{Neutral potassium column density is contingent on
  ionization parameter and fractional depletion of potassium onto 
  grains.  These quantities are inversely proportional to the column 
  density, and therefore the given values above are lower limits.  See
  section 3.4.} 
\tablenotetext{a}{W$_{7665}$ refers to the model fit of the blue line,
in Angstroms.  This measurement is uncorrected for covering factor,
though a correction is included in the column density calculations.}
\end{deluxetable}  



\begin{deluxetable}{lcccc} 
\tablecolumns{5} 
\tablewidth{0pc}
\tablecaption{{\sc Outflow Properties}}
\tablehead{ 
\colhead{Galaxy} & \colhead{v-v$_{sys}$} 
                 & \colhead{R$_{shell}$\tablenotemark{a}}
                 & \colhead{M$_{NaI}$}
                 & \colhead{M$_{HI}$\tablenotemark{b}}\\
\colhead{}       & \colhead{(km s$^{-1}$)}    
                 & \colhead{(pc)} 
                 & \colhead{(M$_\odot$)}
                 & \colhead{(M$_\odot$)}}
\startdata   
NGC 1569    &  -24 &  500  & 36$\pm$1
                            & $4.3\times 10^6\pm 1.2\times 10^5$\\
NGC 1614    & -149 &  1000 & $10^4 \pm$10
                              & $4.2\times 10^8\pm 2\times 10^5$\\
           &  +70 &  1000\tablenotemark{c} 
      & $1.4\times 10^4\pm$10 & $5.8\times 10^8\pm 2\times 10^5$\\
NGC 2363\tablenotemark{d}    
               &  ... & 122  & $<$1.0 & $<2.4\times 10^5$\\
NGC 4214-1\tablenotemark{d}  
             &  ... &  700  &  $<$42  &    $<5\times 10^6$\\
NGC 4214-2  &  -23 &  500  & 127$\pm$16
                             & $1.5\times 10^7 \pm 2\times 10^6$\\
NGC 4449    &  -34 &  1000 & $1.2\times 10^4 \pm$65
                           & 1.4$\times 10^9 \pm 7.8\times 10^6$\\
NGC 5253\tablenotemark{d}
          &  ... &  870  & $<$34    &    $<2.8\times 10^6$\\
M82        &  -91 &  743  & 381$\pm$21  
                            & $1.1\times 10^7\pm 6.5\times 10^5$\\
           &  -35 &  286  & 32$\pm$3.2 
                            & $9.6\times 10^5\pm 9.6\times 10^4$\\
           &  +4  &  500\tablenotemark{c} 
                 & 450$\pm$10 & $1.3\times 10^7\pm 3\times 10^5$\\
           &  +45 &  500\tablenotemark{c} 
                 & 540$\pm$10 & $1.6\times 10^7\pm 3\times 10^5$\\
           &  +86 &  500\tablenotemark{c}   
                 & 140$\pm$10 & $4.2\times 10^6\pm 3\times 10^5$\\
I Zw 18\tablenotemark{d}
              &  ... &  970  & $<$275   &$<4.1\times 10^8$\\
\enddata 
\tablecomments{Masses and errors are computed assuming $f_Df_{ion} =
  100$ and the radii given in the table. Metallicities used are given
  in Table 2. Errors resulting from deviations from the model itself
  (e.g. non-sphericity or other geometrical issues) are not included.}
\tablenotetext{a}{Radii used in superbubble mass calculations are
  referenced in text.} 
\tablenotetext{b}{Atomic hydrogen mass.  Calculated from M$_{NaI}$ as
  discussed in \S4.1.}
\tablenotetext{c}{Radii used for redshifted/infalling components in
  NGC 1614 and M82 are estimates based on the size of the expanding
  superbubbles.}
\tablenotetext{d}{Masses given for NGC 2363, NGC 4214-1, NGC 5253, and
  I Zw 18 are upper limits based on calculations detailed in the
  text and Table 2.} 
\end{deluxetable}  



\clearpage

\begin{thebibliography}{}
\bibitem[Aloisi et al. 2003]{alo03} Aloisi, A., Savaglio, S., Heckman,
  T. M., Hoopes, C. G., Leitherer, C., \& Sembach, K.R. 2003, ApJ, in press
\bibitem[Alonso-Herrero et al. 2001]{ah01} Alonso-Herrero, A.,
  Engelbracht, C. W., Rieke, M. J., Rieke, G. H., \& Quillen, A. C. 2001,
  ApJ, 546, 952
\bibitem[Bajaja et al. 1994]{bajaja} Bajaja, E., Huchtmeier, W. K., \&
  Klein, U. 1994, A\&A, 285, 385
\bibitem[Barlow \& Sargent 1997]{bs97} Barlow, T. A., \& Sargent,
  W. L. W.  1997, AJ, 113, 136
\bibitem[Becker et al. 1995]{beck95} Becker, R., Henkel, C., Bomans,
  D. J., \& Wilson, T. L. 1995, A\&A, 295, 302
\bibitem[Bica et al. 1991]{bica91} Bica, E., Pastoriza, M., Da Silva,
  L., Dottori, H., \& Maia, M.  1991, AJ, 102, 1702
\bibitem[Bushouse 1987]{bus87} Bushouse, H. A. 1987, ApJ, 320, 49
\bibitem[Calzetti et al. 1997]{calz97} Calzetti, D., Meurer, G. R.,
  Bohlin, R. C., Garnett, D. R., Kinney, A. L., Leitherer, C., \&
  Storchi-Bergmann, T. 1997, AJ, 114, 1834
\bibitem[Cameron 1973]{cam73} Cameron, A. G. W. 1973, SSRv, 15, 121
\bibitem[Casoli et al. 1991]{cas91} Casoli, F., Dupraz, C., Combes,
  F., \& Kasez, I., 1991, A\&A, 251, 1
\bibitem[Cervi\~{n}o et al. 1994]{cerv94} Cervi\~{n}o, M., \&
  Mas-Hesse, J. M. 1994, A\&A 284, 789
\bibitem[Dahlem, Weaver, \& Heckman 1998]{dwh98} Dahlem, M., Weaver,
  K. A., \& Heckman, T. M. 1998, ApJS, 118, 401
\bibitem[Gonzalez-Delgado et al. 1998]{gd98} Gonzalez-Delgado, R.,
  Leitherer, C., Heckman, T., Lowenthal, J., Ferguson, H., \& Robert,
  C. 1998, ApJ, 495, 698
\bibitem[Hartmann \& Burton 1997]{hb97} Hartmann, D. \& Burton, W. B.
  1997, Atlas of Galactic Neutral Hydrogen (New York: Cambridge
  University Press) 
\bibitem[Heckman et al. 2000]{heck00} Heckman, T., Lehnert,
  M., Strickland, D., \& Armus, L. 2000, ApJS, 129, 493 (HLSA)
\bibitem[Huchra et al. 1983]{huck83} Huchra, J.P., Geller, M.J.,
  Gallagher, J., Hunter, D., Hartmann, L., Fabbiano, G., \& Aaronson,
  M. 1983, ApJ, 274, 125
\bibitem[Hunter et al. 1998]{hunter1998} Hunter, D. A., Wilcots,
  E. M., van Woerden, H., Gallagher, J. S., \& Kohle, S. 1998, ApJ,
  495, L47 
\bibitem[Jacoby, Hunter \& Christian 1984]{jhc84} Jacoby, G., Hunter,
  D., \& Christian, C. 1984, ApJS, 56, 257
\bibitem[KMH99]{kmh99} Karachentsev, I. D., Makarov, D. I., \&
  Huchtmeier, W. K. 1999, A\&AS, 139,97
\bibitem[Kobulnicky \& Skillman 1995]{ks95} Kobulnicky, H. A., \&
  Skillman, E. D. 1995, ApJ, 454, L121
\bibitem[Kobulnicky \& Skillman 1996]{ks96} ---. 1996, ApJ, 471, 211
\bibitem[Kriss 1994]{krisskross} Kriss, G. 1994, ASP Conf. Ser. 61,
  in Astronomical Data Analysis Software and Systems III,
  ed. D. R. Crabtree, R. J. Hanisch, \& J. Barnes (San Francisco: ASP),
  437 
\bibitem[Kuntz \& Danly 1996]{kd96} Kuntz, K. D. \& Danly, L. 1996,
  ApJ, 457, 703 
\bibitem[Lecavelier des Etangs et al. 2004]{lcde04} Lecavelier des
  Etangs, A., et al. 2004, A\&A, 413, 131
\bibitem[Legrand et al. 2001]{leg01} Legrand, F., Tenorio-Tagle, G.,
  Silich, S., Knuth, D., \& Cervi\~no, M. 2001, ApJ, 560, 630
\bibitem[Lehnert \& Heckman 1996]{lh1996} Lehnert, M. D., \&
  Heckman, T. M. 1996, ApJ, 472, 546
\bibitem[Lehnert, Heckman, \& Weaver 1999]{lhw99} Lehnert, M. D.,
  Heckman, T. M., \& Weaver, K. A. 1999, ApJ, 523, 575
\bibitem[Leitherer et al. 1996]{leith96} Leitherer, C., Vacca, W. D.,
  Conti, P. S., Filippenko, A. V., Carmelle, R., \& Sargent,
  W. L. W. 1996, ApJ, 465, 717
\bibitem[Leitherer et al. 1999]{leith99} Leitherer, C. et al. 1999,
  ApJS, 123, 3
\bibitem[Lequeux et al. 1979]{lequeux79} Lequeux, J., Peimbert, M.,
  Rayo, J. F., Serrano, A., \& Torres-Peimbert, S. 1979, A\&A, 80, 155
\bibitem[Lequeux et al. 1995]{lequeux95} Lequeux, J., Knuth, D.,
  Mas-Hesse, J., \& Sargent, W.  1995, A\&A, 301, 18
\bibitem[Lo et al. 1987]{lo87} Lo, K.-Y., Cheung, K., Masson, C.,
  Phillips, T., Scott, S., \& Woody, D. 1987, ApJ, 312, 574
\bibitem[Luridiana, Peimbert, \& Leitherer 1999]{lpl99} Luridiana, V.,
  Peimbert, M., \& Leitherer, C. 1999, ApJ, 527, 110
\bibitem[MacKenty et al. 2000]{mack00} MacKenty, J. W.,
  Ma\'{i}z-Apell\'{a}niz, J., Pickens, C. E., Norman, C. A., \&
  Walborn, N. R. 2000, AJ, 120, 3007 
\bibitem[Marlowe et al. 1995]{marlowe95} Marlowe, A. T., Heckman, T. M.,
  Wyse, R. F. G., \& Schommer, R. 1995, ApJ, 438, 563
\bibitem[Martin 1996]{crystal96} Martin, C. L. 1996, ApJ, 465, 680
\bibitem[Martin 1998]{crystal98} ---. 1998, ApJ, 506, 222
\bibitem[Martin 1999]{crystal99} ---. 1999, ApJ, 513, 156
\bibitem[Martin \& Armus 2004]{crystal04} Martin, C. L. \& Armus,
  L. 2004, in preparation 
\bibitem[Martin \& Zalubas 1981]{mz81} Martin, W. C. \& Zalubas,
  R. 1981, J. Phys. Chem. Ref. Data, 10, 153
\bibitem[Martin, Kobulnicky, \& Heckman(2002)]{2002ApJ...574..663M}
  Martin, C. L., Kobulnicky, H. A., \& Heckman, T. M. 2002, \apj, 574, 663
\bibitem[McIntyre 1998]{1998PASA...15..157M} McIntyre, V. J.\ 1998,
  Publications of the Astronomical Society of Australia, 15, 157  
\bibitem[Meier, Turner, Crosthwaite, \& Beck 2001]{mtcb01} Meier,
  D. S., Turner, J. L., Crosthwaite, L. P., \& Beck, S. C. 2001, \aj, 121,
  740  
\bibitem[Morton 1991]{morton91} Morton, D. C. 1991, ApJS, 77, 119
\bibitem[M{\" u}hle, H{\" u}ttemeister, Klein, \& 
Wilcots 2001]{2001AGM....18S0546M} M{\"u}hle, S., H{\"u}ttemeister,
  S., Klein, U., \& Wilcots, E. M.\ 2001, Astronomische Gesellschaft
  Meeting Abstracts, 18, 546 
\bibitem[Neff et al. 1990]{neff90} Neff, S. G., Hutchins, J. B.,
  Standord, S. A., \& Unger, S. W. 1990, AJ, 99, 1088
\bibitem[Ohyama et al. 2002]{ohyama} Ohyama, Y., et al. 2002, in ASP
  Conf. Ser. 289, The Proceedings of the IAU 8th Asian-Pacific
  Regional Meeting, ed. S. Ikeuchi, J. Hearnshaw, \& T. Hanawa (San
  Francisco: ASP), 285
\bibitem[Papaderos et al. 1996]{pap97}Papaderos, P., Loose,
  H.-H. Thuan, T. X., \& Fricke, K. J.,1996, A\&AS, 120, 207
\bibitem[Phillips 1993]{phil93} Phillips, A. 1993, AJ, 105, 486
\bibitem[Roy et al. 1991]{roy91} Roy, J.-R., Boulesteix, J., Joncas,
  G., \& Grundseth, B. 1991, ApJ, 367, 141
\bibitem[Rupke, Veilleux, \& Sanders 2002]{rvs02} Rupke, D. S.,
  Veilleux, S., \& Sanders, D. B. 2002, ApJ, 570, 588
\bibitem[Sargent \& Fillipenko 1991]{sf91} Sargent, W. L. W., \&
  Fillipenko, A. V. 1991, AJ, 102, 107
\bibitem[Savage \& Sembach 1991]{ss91} Savage, B. D., \& Sembach,
  K. R. 1991, ApJ, 379, 245
\bibitem[Savage \& Sembach 1996]{ss96} ---. 1996, ARA\&A, 34, 279 
\bibitem[Silich \& Tenorio-Tagle 1998]{sil98} Silich, S. A., \&
  Tenorio-Tagle, G. 1998, MNRAS, 299, 249
\bibitem[Skillman, Kennicutt, \& Hodge 1989]{skh89} Skillman, E. D.,
  Kennicutt, R. C., \& Hodge, P. W. 1989, ApJ, 347, 875
\bibitem[Sofue 1997]{sofue} Sofue, Y. 1997, PASJ, 49, 17
\bibitem[Spitzer 1968]{1968dms..book.....S} Spitzer, L. 1968, Diffuse
  Matter in Space (New York: Interscience Publication)
\bibitem[Spitzer 1978]{1978ppim.book.....S} ---. 1978, Physical
  Processes in the Interstellar Medium (New York: Wiley-Interscience)
\bibitem[Spitzer \& Fitzpatrick 1995]{sfitz95} Spitzer, L. \&
  Fitzpatrick, E. L. 1995, ApJ, 445, 196
\bibitem[Stil \& Israel2002]{2002A&A...392..473S} Stil, J. M., \&
  Israel, F. P. 2002, \aap, 392, 473 
\bibitem[Storchi-Bergmann \& al. 1994]{sb94} Storchi-Bergmann, T.,
  Calzetti, D., Kinney, A. L. 1994, ApJ, 429, 572
\bibitem[Summers et al. 2003]{summers03} Summers, L. K., Stevens,
  I. R., Strickland, D. K., \& Heckman, T. M. 2003, MNRAS, 342, 690
\bibitem[Umeda et al. 2002]{umeda02} Umeda, H., Nomoto, K., Tsuru,
  T. G., \& Matsumoto, H. 2002, ApJ, 578, 855
\bibitem[van Zee et al. 1998]{leise98} van Zee, L., Westpfahl, D.,
  Haynes, M. P., \& Salzer, J. J. 1998, AJ, 115, 1000
\bibitem[Vogt et al. 1994]{vogt94} Vogt, S. S., et al. 1994,
  Proc. SPIE, 2198, 362 
\bibitem[Wakker \& Mathis 2000]{wm00} Wakker, B. P. \& Mathis, J. S. 2000,
  ApJ, 544, L107
\bibitem[Wei\ss~et al. 1999]{weiss99} Wei{\ss}, A., Walter, F.,
  Neininger, N., \& Klein, U. 1999, A\&A, 345, L23
\bibitem[Wills, Pedlar, \& Muxlow 2002]{2002MNRAS.331..313W} Wills,
  K. A., Pedlar, A., \& Muxlow, T. W. B. 2002, \mnras, 331, 313 
\end{thebibliography}
\end{document}